\documentclass[
reprint,superscriptaddress,amsmath,amssymb,aps,pra]{revtex4-1}
\usepackage{color}
\usepackage{graphicx}% Include figure files
\usepackage{caption}
\usepackage{subcaption}
\usepackage{dcolumn}% Align table columns on decimal point
\usepackage{bm}% bold math
\usepackage{braket}% braket quantum symbols
\usepackage{float}% fix figure position
\usepackage{multirow}
\usepackage{hhline}
\usepackage{mathtools}
\usepackage{qcircuit}
\usepackage{lipsum}

\begin{document}
\title{Optimized Measurement Schedules for the Surface Code with Dropout}
\author{Benjamin Anker}
\email{banker@unm.edu}
\affiliation{Google Quantum AI, Venice, CA 90291}
\affiliation{Center for Quantum Information and Control (CQuIC), University of New Mexico}
\author{Dripto M. Debroy}
\email{dripto@google.com}
\affiliation{Google Quantum AI, Venice, CA 90291}
\begin{abstract}
Recent work has shown that fabrication defects can be well-handled using a strategy relying on the mid-error-correction-cycle state~\cite{debroy2024lucisurfacecodedropouts}.
In this work we present two improvements to the original prescription.
First, we quantify the impact of the choice of a more complete set of gauge operators originally proposed for the hex-grid surface code~\cite{higgott2025handlingfabricationdefectshexgrid} on the standard square-grid surface code, as well as a new method for excising effectively unused qubits.
Second, we leverage the expressivity of the LUCI framework as an intermediate representation, using integer linear programming to find performant physical circuits from the large space of valid LUCI circuits.
We show that on the $d = 11$ surface code at $1\%(3\%)$ dropout rate for qubits and couplers, these optimizations allow for a total improvement of $14.5\%(23.6\%)$ over $4d$ round of syndrome extraction using the SI1000 noise model at $0.1\%$ noise~\cite{gidney2021honeycomb}.
\end{abstract}
\maketitle
\section{Introduction}\label{sec:intro}
In order to achieve the reliable, large-scale quantum computing necessary for useful quantum algorithms, quantum error correction is essential.
Concerted effort over the past several years~\cite{krinner2022realizing,google2023supressing,google2024threshold} has shown that the best modern hardware is capable of enabling arbitrarily scalable quantum memory, as long as larger devices can continue to be built.
These demonstrations have used the surface code implemented on superconducting hardware~\cite{krantz2019quantum} using a planar architecture.
As we scale chip sizes, it is unrealistic to assume that every component works reliably -- fabrication errors cause some components to be unavailable, while transient errors can render them unavailable for significant time periods.
We term this unavailability as dropout.

Ideally, it should be possible to account for moderate dropout rates by routing information around defects, rather than simply discarding an entire chip.
%TODO: mention SnL explicitly
Indeed, there are a variety of known solutions to this problem~\cite{wei2024lowoverheaddefectadaptivesurfacecode,nagayama2017surface,auger2017fault,wei2024lowoverheaddefectadaptivesurfacecode,GransSamuelsson2024improvedpairwise,leroux2024snakesladdersadaptingsurface,mishmash2025excisingdeadcomponentssurface}.
In this work, we consider the LUCI framework~\cite{debroy2024lucisurfacecodedropouts}, a state-of-the-art method for handling dropout which is based upon the mid-cycle state of the surface code~\cite{mcewen2023relaxing}.
The motivating insight of the LUCI framework is that since the mid-cycle state of the rotated surface code is a larger, non-rotated surface code supported on both data and measure qubits, dropout on both kinds of qubits can be treated symmetrically.

As originally presented, LUCI was effectively a single-point design, i.e. the original algorithm to construct a LUCI diagram defined exactly one circuit for a given dropout configuration.
Here, we more fully appreciate the fact that the framework can instead be seen as an intermediate representation (IR) -- a flexible language defining an entire space of valid circuits, rather than just a single instance.

First, we establish a stronger baseline by quantifying the improvement to logical error rate (LER) provided by defining a LUCI diagram with respect to the more complete set of gauge operators originally presented~\cite{higgott2025handlingfabricationdefectshexgrid} for the hex-grid surface code, as well as removing some redundant stabilizer operators and the qubits they are supported on.
Second, as the main focus of this work, we introduce a new optimization pass using integer linear programming that operates on the IR defined by LUCI diagrams.
We introduce a proxy for LER as a (linear) objective in terms of our IR and demonstrate how to formally optimize the circuit structure.
We show that for the case we consider, the $d = 11$ surface code, the new choice of gauge group improves the LER by $8.2\%(14.9\%)$ at $1\%(3\%)$ equal qubit and coupler dropout rates, while optimizing the circuit structure provides a \emph{further} $6.8\%(10.3\%)$ reduction to LER, demonstrating the significant gains possible with this formal approach.
%TODO: adjust if we do a little hyperparameter tuning
It is notable that we achieve these gains without relying on expensive hyperparameter tuning, although investing the time to do so in the future could improve the numerical impact of our work.

We note concurrent work by Wolanski~\cite{wolanski2025automatedcompilationincludingdropouts} (ACID) which utilizes similar optimization methods to maximize measurement density.
While high measurement density is a desirable heuristic, we did not observe a direct correlation with improved LER in simulations using our methods (illustrated in Figures~\ref{fig:schedule_comparison} and~\ref{fig:max_vs_orig}). 
In Appendix~\ref{appendix:acid} we directly compare the performance of ACID versus the canonical implementation of LUCI, finding that LUCI produces significantly lower LER.

Section~\ref{sec:background} reviews the original LUCI framework.
In Section~\ref{sec:wt-one-gauges} we quantify the impact of modifying the gauge group with respect to some weight-one operators.
Section~\ref{sec:ilp-model} casts the minimization of LER by deciding upon a LUCI diagram as a discrete optimization problem, detailing a justification for each term in the objective function we use as a proxy for LER.
In Section~\ref{sec:ilp-data} we present the numerical results of optimizing the proposed model as an integer linear programming (ILP)~\cite{wolsey1998ip} problem, demonstrating the utility of the approach.
Finally in Section~\ref{sec:conclusion} we discuss the limitations of our work and avenues for future improvement.
\section{The LUCI framework}\label{sec:background}
Before describing the numerical experiments and advances we have made, we first summarize the LUCI framework.
As it is commonly presented, the (rotated) surface code consists of a set of weight-four stabilizer operators arranged in a checkerboard pattern, with weight-two stabilizers on the boundaries.
Measuring these weight-four stabilizers can be done in a distance-preserving manner with a single syndrome qubit by recognizing that weight-one errors which propagate to a weight-two error are harmless, as long as the weight-two error is perpendicular to the logical operator of the same Pauli type.
In most architectures, the single syndrome qubit associated to each stabilizer lies at the center of the stabilizer it is used to measure.

\begin{figure}
    \includegraphics[width=\linewidth]{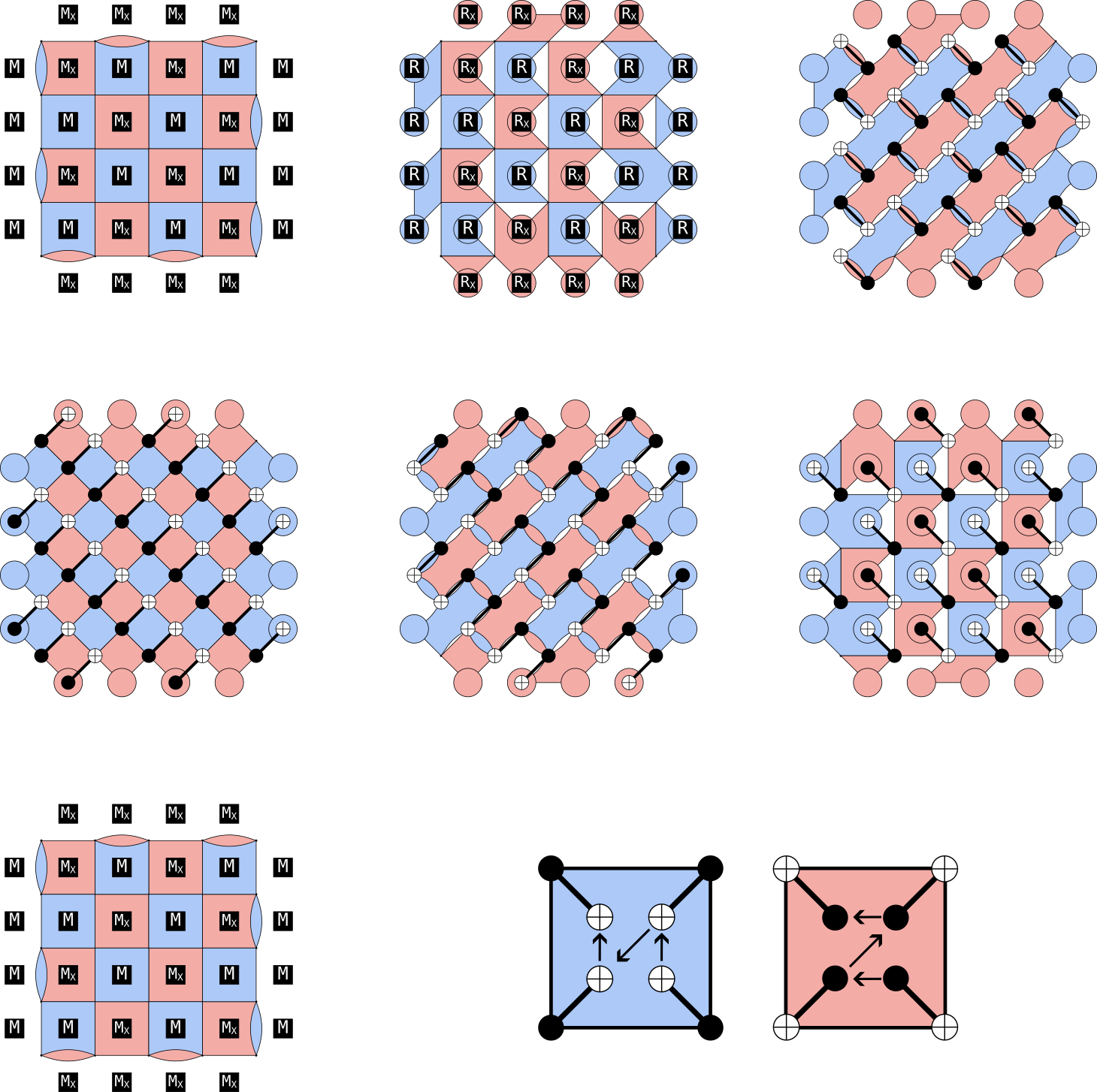}
    \caption{\label{fig:rotated_to_unrotated}Here we track each stabilizer of the rotated surface code plus syndrome qubit system. 
    We see that after the first two layers of \textsc{cnot}s we land in the unrotated surface code state before returning to the rotated surface code state. 
    Note that the familiar stabilizers we end up with are actually the result of evolving the single-qubit stabilizers.
    We choose a basis of weight-five and weight-one stabilizers instead of weight-one and weight-four to make the transformation more obvious.}
\end{figure}

As stated, however, this description does not consider the stabilizers of the syndrome qubits.
It is obvious that one can define these stabilizers, the stabilizer associated with a syndrome qubit prepared in the $\ket 0$ state being the $Z$ operator and similarly for $\ket +$ and $X$, but it is not obvious why one would want to.
Explicitly considering them as part of the stabilizer group and tracking the evolution of the stabilizer group through the first two \textsc{cnot} layers of the canonical syndrome extraction circuit yields an unrotated surface code supported on both data and syndrome qubits as seen in Figure~\ref{fig:rotated_to_unrotated}~\cite{mcewen2023relaxing}.

\begin{figure}[h!]
    \includegraphics[width=\linewidth]{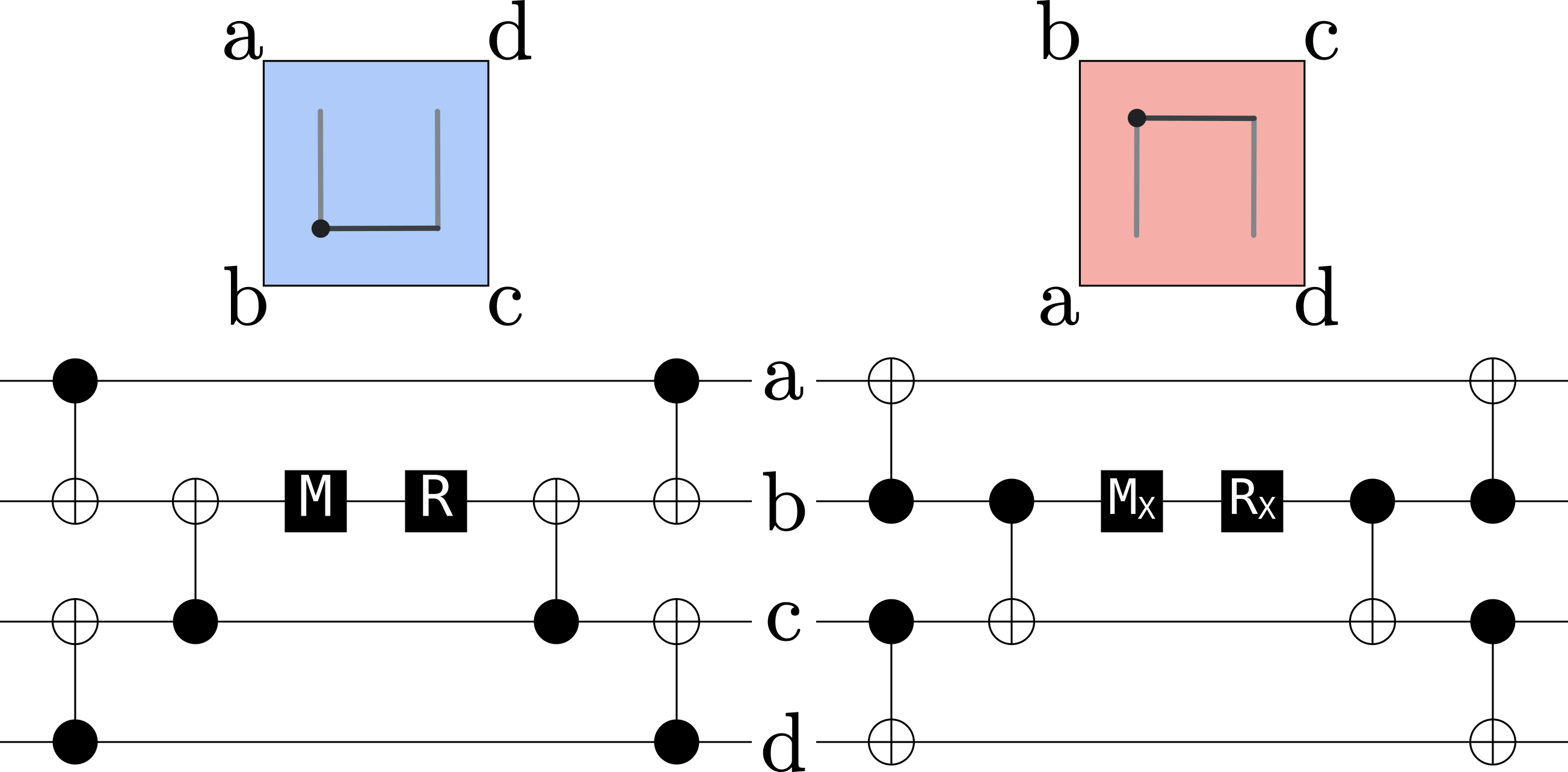}
    \caption{\label{fig:plaquette}An example of a $Z$-type and $X$-type shape.
    The light gray legs of the shape correspond to the first \textsc{cnot} layer, the dark gray crossbeam corresponds to the second layer, and the dot corresponds to the measurement.
    This circuit is then repeated in reverse (substituting a reset for the measurement) to create the new weight-four stabilizer.
    For the sake of clarity we have labeled the qubits in this example.}
\end{figure}

Using this observation, we can trivially redefine repeated rounds of syndrome extraction.
Instead of considering rounds $r_1, r_2, r_3, \ldots$, we consider rounds $r_1^a, r_1^br_2^a, r_2^br_3^a, r_3^br_4^a, \ldots$, where $r_i^a$ is composed of the $i-$th reset layer together with the first two \textsc{cnot}s of the $i$-th layer, and $r_i^b$ is composed of the final two \textsc{cnot}s together with the measurement of the syndrome qubit.
In this setting, each of the (twice as many) stabilizers of the mid-cycle state is measured once every other round -- $r_i^b$ shrinks a stabilizer to a single qubit and measures it out, while $r_{i + 1}^a$ grows the stabilizer of the reset qubit back to where it started.
In round $r_{i + 1}^b$ this stabilizer grows instead of shrinking, is not measured out, and then shrinks back to a weight-four.
We can use this mid-cycle state as the state defining our code, with the understanding that we apply a half-round to come from or return to the standard rotated surface code state.
In this picture we represent the circuit used to measure each stabilizer operator using a compact graphical depiction, illustrated in Figure~\ref{fig:plaquette}.
These subcircuits, or shapes, (two layers of \textsc{cnot}s, a measurement, a reset, and another two layers of \textsc{cnot}s), each measure and re-prepare a stabilizer or gauge operator, and are the intermediate representation we optimize over in Section~\ref{sec:ilp-model}.

The benefit of this picture is that it removes any distinction between data and measure qubits.
This allows us to easily define a strategy to handle dropout which can handle data qubit dropout, measure qubit dropout, and coupler dropout in a relatively symmetrical manner.
Before we outline the full picture we must decide on which gauge operators we will measure.

\begin{figure*}[ht!]
    \includegraphics[width=\textwidth]{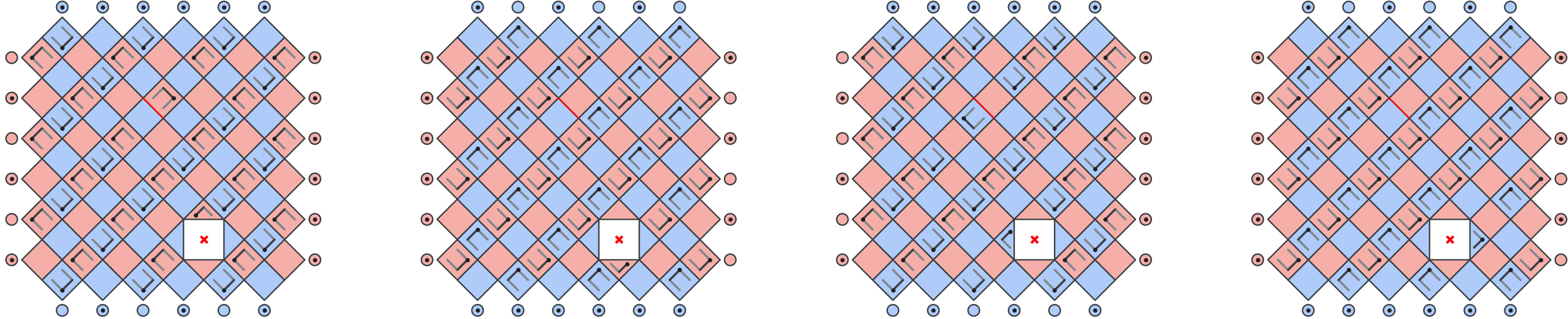}
    \caption{\label{fig:luci_example}Here we provide an example of how a missing qubit and missing coupler are treated using a LUCI diagram. This diagram corresponds to four mid-cycle to mid-cycle rounds.}
\end{figure*}

The algorithm that defines the gauge operators is simple.
First, any broken qubits are removed from the support of the (mid-cycle) stabilizers with support on them, most likely turning some stabilizers into gauge operators.
Then each operator, whether gauge or stabilizer, is split into connected components of qubits, with connections being couplers it is possible to use.
Finally, if this procedure has split the surface code patch into two or more entirely disconnected regions, the largest is kept while the rest are discarded.
The resulting set of operators will be measured out.
This algorithm is slightly simpler than the original LUCI algorithm, since it implicitly allows for the use of weight-one gauge operators rather than requiring each pair of orthogonal couplers to be converted into a broken qubit and each broken qubit to be converted into four broken couplers.
The intuition behind the utility of the weight-one gauge operators, beyond just making the algorithm simpler to describe, is outlined in Section~\ref{sec:wt-one-gauges}.

We now can describe the shapes we use to measure these gauge operators, in place of the stabilizer operators.
Effectively, we take the shape used to measure the entire stabilizer, rotate it until it does not use any broken couplers, then remove any portion of it with support on a broken qubit.
Since the gauge operators we choose are guaranteed to be connected by non-broken couplers this procedure is always possible, but can cause incompatibilities with neighboring shapes.
To ensure each operator is measured at least once, we four-color the set of operators measured and give each color priority in one round.
This four-round measurement schedule is summarized and depicted graphically by four boards, illustrated on the distance-$7$ surface code in Figure~\ref{fig:luci_example}.

Although the LUCI framework functions according to the mid-cycle-state, where data and measure qubits are treated symmetrically, this is not always experimentally convenient.
To avoid adding any restrictions on when this work is applicable, we restrict ourselves to only measuring the canonical measure qubits (i.e. syndrome qubits with respect to the rotated surface code).
This means that we may have to adjust the gauge operators decided upon, removing any data qubit which supports a weight-one gauge operator and recomputing.
It also means that the algorithm for picking shapes must be aware of the distinction between data and measure qubits.

\section{Weight-one Gauge and Stabilizer Operators}\label{sec:wt-one-gauges}
It has already been shown~\cite{higgott2025handlingfabricationdefectshexgrid} that the original LUCI prescription at times failed to effectively make use of all non-broken hardware.
This was particularly troublesome when considering a hex-grid surface code layout, since in this picture one quarter of the couplers are effectively already broken.
The fact that LUCI did not handle two perpendicular broken couplers as efficiently as possible made a single (extra) missing coupler irredeemably ruin the resulting distance on the hex-grid.
We briefly summarize how allowing for weight-one gauges can allow for better performance, but refer the reader to previous work~\cite{higgott2025handlingfabricationdefectshexgrid} for more details.

\begin{figure}[h!]
    \includegraphics[width=0.75\linewidth]{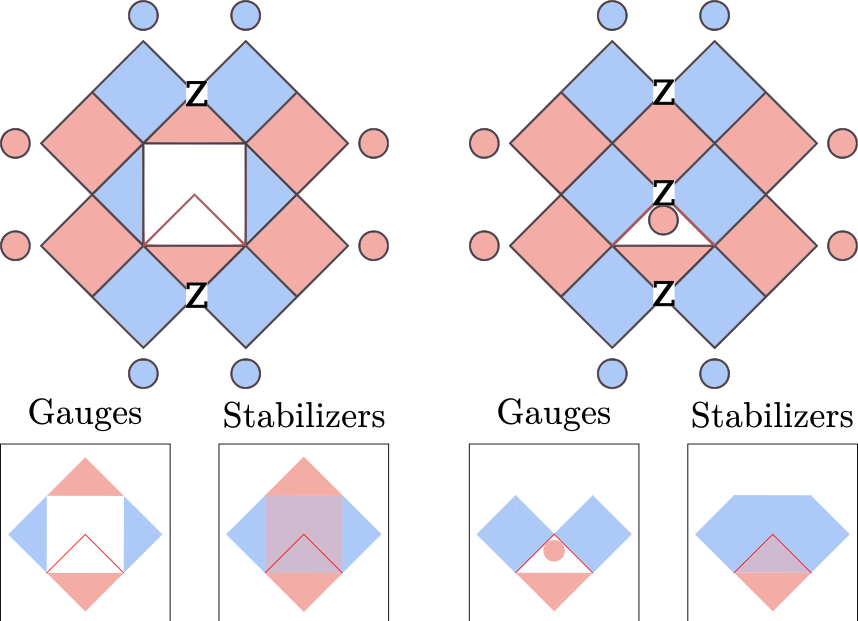}
    \caption{\label{fig:wt_one_gauge_impact}
    The stabilizers produced for a fixed dropout configuration when we disallow weight-one gauge operators as in the original prescription~\cite{debroy2024lucisurfacecodedropouts} (left) versus when we allow weight-one gauge operators as in more recent work~\cite{higgott2025handlingfabricationdefectshexgrid}. 
    One minimum weight representative of logical $Z$ is marked in each case.}
\end{figure}

Examining the super-stabilizers produced by a set of gauges which include a weight-one gauge, as in the bottom right of Figure~\ref{fig:wt_one_gauge_impact}, we see that the two weight-four $Z$ operators which in the absence of dropout would be stabilizers are actually gauge operators.
The super-stabilizer they multiply to form has no support on the qubit with the weight-one gauge operator.
In this sense, it seems that we have completely removed the physical qubit from the code, and that it cannot help us do error correction.

However, if we consider the opposite basis, the picture clarifies.
We see on the left in the case where we do not include a weight-one gauge operator, there exists a weight-two logical operator (marked with $Z$s on the figure).
When we add the weight-one gauge operator, it is not possible to find a weight-two logical operator composed of $Z$s -- the $Z$ distance increases to $3$.

We also make a second optimization to the structure of the mid-cycle stabilizers, this time by \emph{removing} certain weight-one operators.
The qubits that we remove are (canonical syndrome) qubits on the boundary which are only measured as part of a weight-one operator and never used by a measurement subcircuit as a `waypoint' qubit between the measurement qubit and another qubit in the support of the same operator.
Before removal these qubits support a weight-one stabilizer operator.

In the original construction, such boundary qubits were always kept so that they could be used to route around dropout.
\begin{figure}[h!]
    \centering
    \begin{subfigure}[b]{\linewidth}
        \centering
        \includegraphics[width=0.88\linewidth]{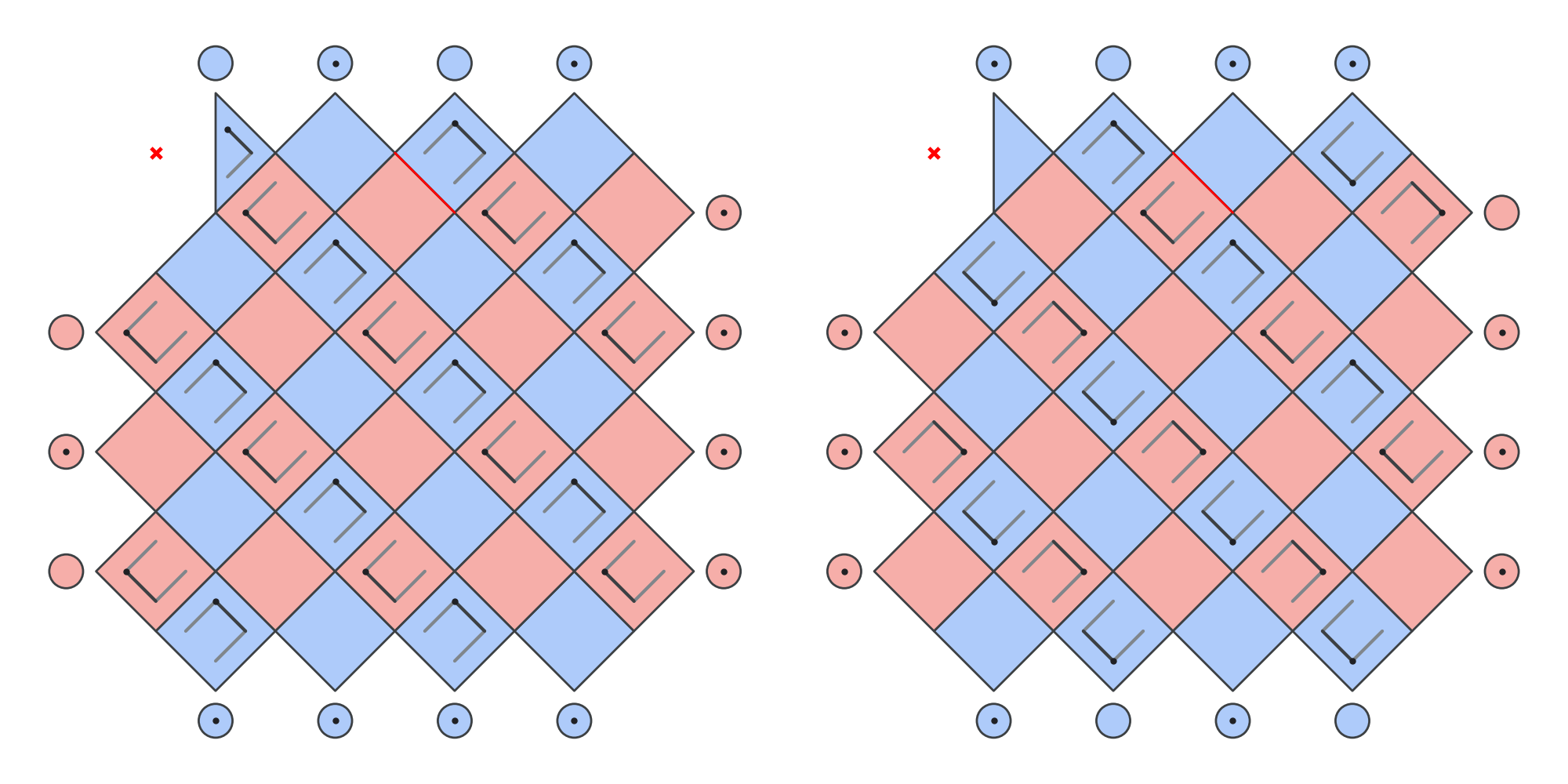} 
        \caption{\label{fig:no_removal}Without removing unnecessary boundary qubits.}
    \end{subfigure}
    \begin{subfigure}[b]{\linewidth}
        \centering
        \includegraphics[width=0.88\linewidth]{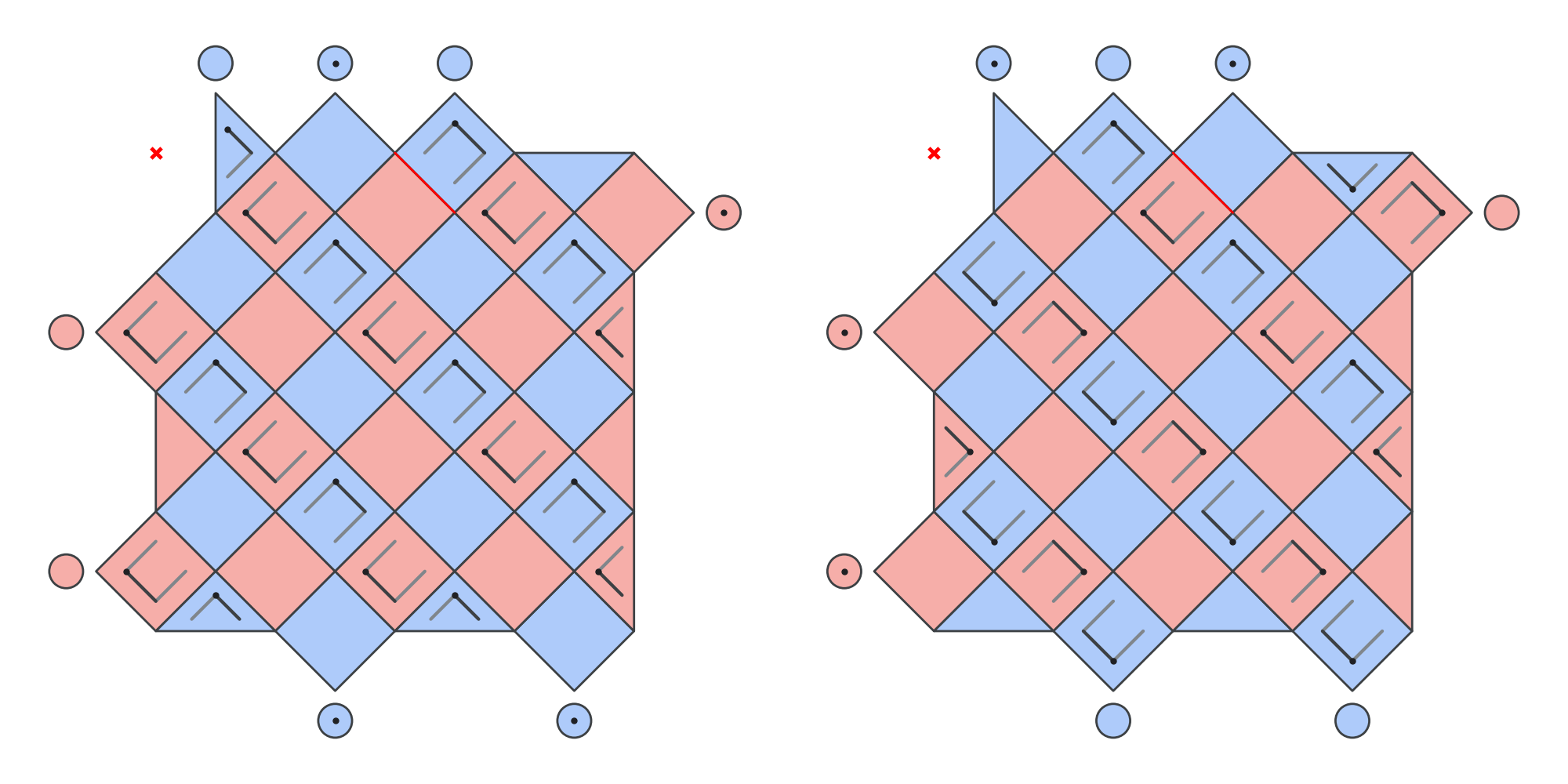}
        \caption{\label{fig:removal}Removing unnecessary boundary qubits.}
    \end{subfigure}
    \caption{Comparison of the same two-round solution for a given dropout configuration with and without removing unnecessary boundary qubits.
    The boundaries in~\ref{fig:removal} are asymmetrical because the usage of the syndrome qubits on the boundary in~\ref{fig:no_removal} is asymmetrical.}
    \label{fig:weight_one_removal}
\end{figure}
This optimization removes those that end up not being used for this purpose.
In Figure~\ref{fig:weight_one_removal} we compare the mid-cycle states before and after trimming off a number of weight-one stabilizers.

We should expect that these removed qubits are never helpful because
\begin{itemize}
    \item The extra flexibility they afford for choosing a measurement schedule is never used (since they are never a waypoint or measure qubit)
    \item They are stabilizers, not gauge operators, meaning their measurement is not used to form a superstabilizer but only to protect the very qubit they sit on
    \item They increase the size of detectors on the boundaries, and create a small detector in the support of a larger one.
\end{itemize}
In specifically chosen examples like those in Figure~\ref{fig:wt_one_gauge_impact}, or in a hex-grid architecture where LUCI as originally formulated fails, it is also clear that including weight-one gauges should yield a large benefit.

It is not entirely clear though how much one should expect the respective exclusion or inclusion of these weight-one stabilizer or gauge operators to help in the average case on a square-grid architecture.
We examine this question briefly, in order to provide a baseline to compare against when assessing the impact of the optimization we perform in Sections~\ref{sec:ilp-model} and~\ref{sec:ilp-data}.
In Figure~\ref{fig:wt-one-data} we summarize data showing an improvement in LER of $8.2\%(14.9\%)$ for $1\%(3\%)$ dropout.

Simulation results were obtained using Stim~\cite{gidney2021stim} and a matching based decoder~\cite{Higgott2025sparseblossom}, taking $30$ million shots for $1\%$ and $10$ million for $3\%$ (the increased shots in the $1\%$ case necessary to estimate the smaller LER rates with similar confidence intervals).
Circuits consist of $4d$ rounds of syndrome extraction, so that the bulk time-like region dominates the LER behavior.
\section{LUCI Diagrams as an Optimization Problem}\label{sec:ilp-model}
Before presenting the model we optimize and the results we obtain we first provide intuition on the features of the heuristic schedule we wish to improve on.
\begin{figure*}
    \includegraphics[width=0.8\linewidth]{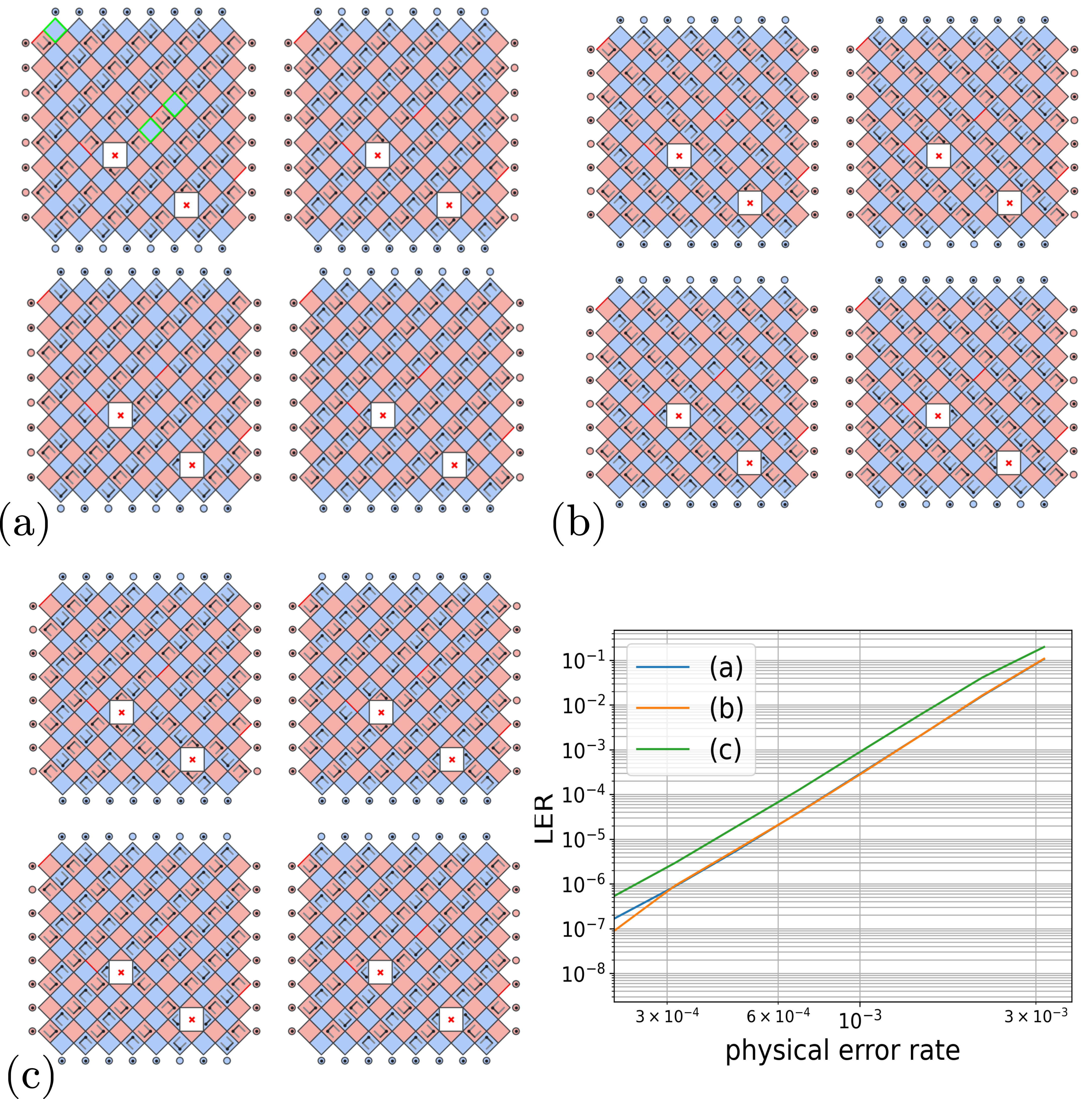}
    \caption{Comparison of measurement schedules for a fixed dropout configuration (each $2\times2$ grid representing one measurement schedule). 
    Top left we see an example of how some configurations can interact poorly with the algorithm outlined in Section~\ref{sec:background}.
    Green-boxed squares in the first round are not measured, with similar measurement gaps in following rounds.
    Top right we mitigate the issue with simple heuristics, flipping and rotating diagonals without fundamentally changing the algorithm, but do not observe a decreased LER (bottom right).
    A na\"ive maximal schedule (bottom left) fills more gaps but results in higher LER, highlighting the need for the more sophisticated metrics employed in Section~\ref{sec:ilp-model}.
    }
    \label{fig:schedule_comparison}
\end{figure*}
Consider Figure~\ref{fig:schedule_comparison}.
In this example, we see the shortcomings of the original algorithm, in which some stabilizers are only measured once every four rounds because of incompatibilities between the default orientation of the LUCI shape used to measure them and the LUCI shape that dropout requires for neighboring squares.
This reduction in measurement frequency means that errors that the measurement of this stabilizer should detect have longer to build up, increasing the logical error rate.

In this case a denser measurement schedule is achievable simply by vertically reflecting each of the LUCI shapes asked for by the default algorithm.
This happens to allow several more measurements to fit.
In general, though, simple algorithms such as ``flip each row to maximize the number of measurements'' do not improve logical error rate.
For example, the maximally dense schedule given in Figure~\ref{fig:schedule_comparison} produces a LER $5.5\times$ the LER produced by the default schedule over $4d$ rounds at $0.1\%$ noise, despite having $15$ more measurements per four-round cycle, while a simple heuristic algorithm adds $5$ measurements but does not change logical error rate at all.

We examine the quantitative impact of maximizing measurement in more detail in Section~\ref{sec:ilp-data} and Appendix~\ref{appendix:acid}, but we can qualitatively explain this difference just by considering the space-time volume covered by each detecting region, which we term detector volume.
Although it is true that skipping a stabilizer measurement increases the volume of the detector defined by consecutive measurements of this stabilizer, this is not the only way to increase its volume.
We return to this perspective in Appendix~\ref{appendix:max_meas_bad}, examining the structure of the circuits given as an example in Figure~\ref{fig:schedule_comparison}.
The choice of neighboring shapes can also increase the number of errors this detector is sensitive to, a fact which we return to in Section~\ref{sec:proxy}.

The interplay between these terms is non-trivial, and it is not immediately clear what impact one will have relative to another.
In a sense, the motivation of this work is to reduce the problem to a model in which we can precisely quantify the relative importance of each term.
In the remainder of this section we do exactly this, formulating the problem of choosing a LUCI diagram as an integer linear program (ILP)~\cite{wolsey1998ip} which we optimize using the ideas we have just sketched.

There are (at least) two ways to understand why larger detector volumes might be harmful in terms of LER.
The first is that the violation of a large detecting region gives less information than the violation of a small detecting region, in the sense that there are more consistent errors when the large region is violated.
The second is that a large detecting region permits the existence of more minimum weight logical operators (which one can observe by considering the graph in which qubits share an edge if they both lie in the support of the same detecting region, paths through such a graph describing all of the undetectable errors).
In either picture, logical error rate is minimized by equalizing detector volumes as much as possible, rather than decreasing one's volume at the expense of another.

\subsection{ILP Formulation}
Here we describe in detail each feature of the model we optimize over to obtain the results in Section~\ref{sec:ilp-data}.
We represent a LUCI diagram as a set of boolean variables $v_{t, i, p}$, where $t$ is the board index (i.e. time-slice), $i$ is the operator index (each gauge or stabilizer operator being assigned a consistent index across time), and $p$ is the shape used to measure operator $i$ at time $t$.
Note that for each $(t, i)$ there are several boolean variables $v_{t, i, p}$, each one corresponding to one shape that could measure operator $i$.
Exactly how many of these variables there are depends on the dropout configuration, but for a weight-$4$ operator there are $8$ shapes which can measure it (4 directions for a shape to face, with a choice of two measure qubits in each case), for a weight-$3$ there are $4$ ($3$ choices of measure qubit, with exactly one choice for all but one of them), for a weight-$2$ there are also $4$ (a choice of measure qubit, and a choice of whether to put the \textsc{cnot} in the first or second layer), and for a weight-$1$ there is one.
As noted in Section~\ref{sec:background}, however, we restrict ourselves to only measuring canonical data qubits, reducing the number of possibilities for each operator by a factor of approximately two.

\begin{figure}
    \includegraphics[width=0.8\linewidth]{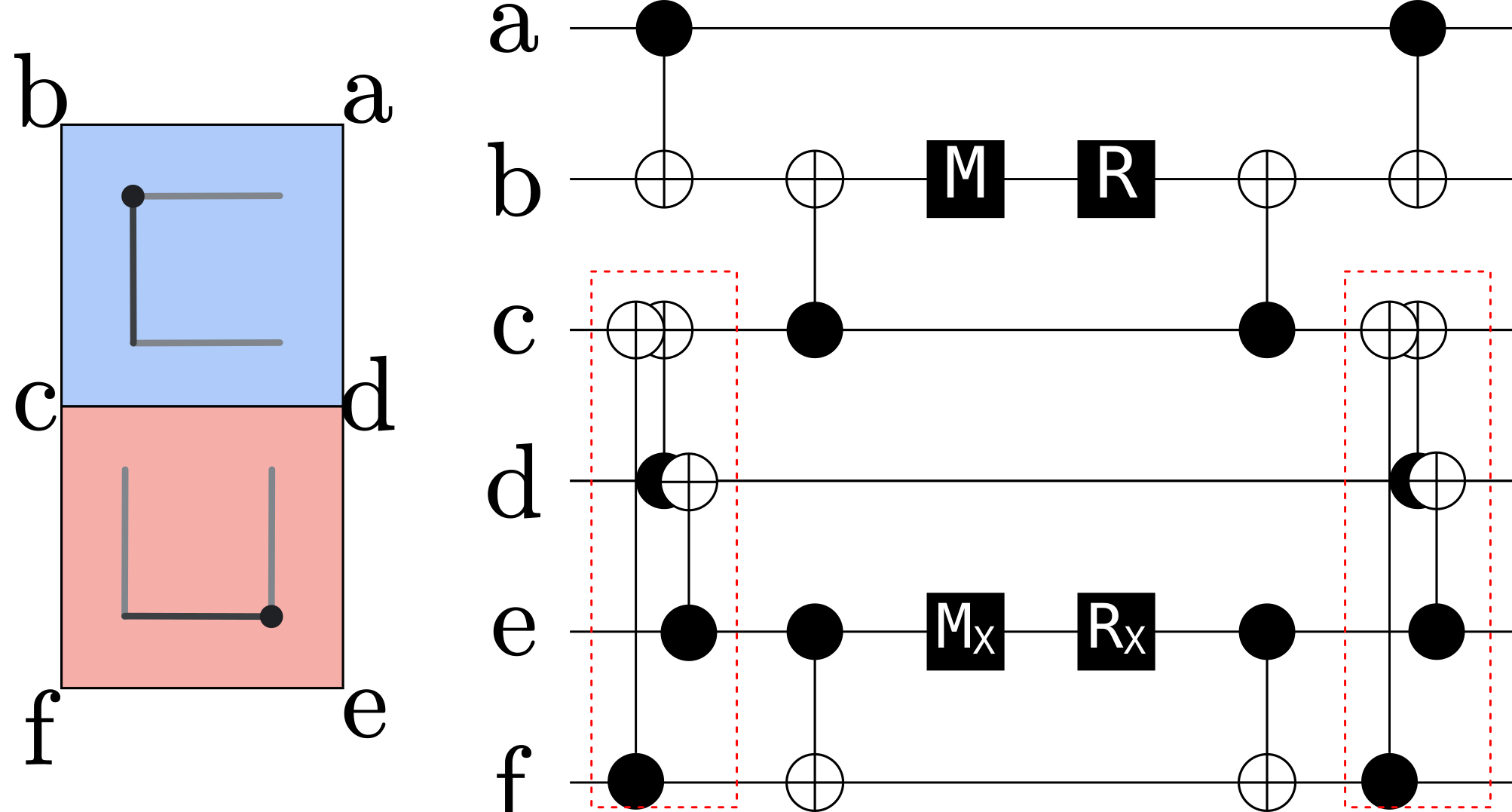}
    \caption{\label{fig:incompatibility_example}It is unphysical to ask for the two shapes on the left to be simultaneously implemented, since their implementations use the same qubit at the same time for different purposes (red dashed boxes).}
\end{figure}

On these variables we put a variety of constraints.
The first constraint encodes the fact that some subcircuits are physically incompatible with each other, as in Figure~\ref{fig:incompatibility_example}.
To ensure that the diagram the optimizer creates is physically meaningful, we demand
\[
    \forall_t : v_{t, i, p} + v_{t, j, q} \leq 2 - I_{i, p, j, q}
\]
where $I_{i, p, j, q}$ is the indicator function encoding whether using shape $p$ to measure operator $i$ is incompatible with simultaneously using shape $q$ to measure operator $j$.
More precisely, two shapes measuring $o_i, o_j$ are compatible only if they do not ask for multiple \textsc{cnot}s to use the same qubit at the same time, $[o_i, o_j] = 0$, and if implementing them simultaneously correctly measures both $o_i$ and $o_j$.

Secondly, we require
\begin{align}
    \forall_i :& \sum_t \sum_p v_{t, i, p} &\geq 1\label{eqn:measure_once}\\
    \forall_{i, t} :& \sum_p v_{t, i, p} &\leq 1
\end{align}
This encodes the requirement that each operator is measured at least once, and the fact that it is not physically meaningful to ask for an operator to be simultaneously measured in two different ways.

Finally, we put constraints on the model designed to ensure that each superstabilizer is measured at some point.
It is easiest to describe this using the auxiliary variable
\[
    f_{t, i} := \bigvee_p v_{t, i, p},
\]
which provides a convenient way to ask whether operator $i$ is measured at time-slice $t$ without specifying \emph{how} it is measured.
Implementing this variable requires appropriate linearization.
We require that for any superstabilizer composed of the product of (gauge) operators $i_1, \ldots, i_k$ we have
\begin{equation}\label{eqn:superstab}
    \exists t : \bigwedge_{j = 1}^k f_{t, i_j} \lor f_{t + 1, i_j};
\end{equation}
that is, for each superstabilizer there exists a pair of consecutive rounds such that each gauge operator composing the superstabilizer is measured in one of these two rounds.
This ensures that the eigenvalue of each superstabilizer can actually be inferred at some point~\cite{suchara2011subsystem}, but is a sufficient condition rather than a necessary one -- it is only used because it is relatively simple.
This set of constraints must also be linearized.

For all of the definitions and constraints, including those in the following sections, we linearize logical AND and OR by introducing an auxiliary variable $y$ to hold the value of AND$(a, b)$ or OR$(a, b)$ by imposing the following conditions:
\begin{align*}
    \text{AND}(a, b) = \begin{cases}
        y \leq a\\
        y \leq b\\
        y \geq a + b - 1
    \end{cases}\\
    \text{OR}(a, b) = \begin{cases}
        y \geq a\\
        y \geq b\\
        y \leq a + b\text{\ \ \ \ \ }
    \end{cases}
\end{align*}
Constraining the value of $y$ then constrains the logical expression we are actually interested in.
This can easily be generalized to an arbitrary number of terms using a big-M strategy.
We should also note that the existential quantifier $\exists$ is effectively logical OR.
Although logically the universal quantifier $\forall$ is equivalent to logical AND, we do not have to linearize the terms in this section using the quantifier, since we can just add each term independently.

This relatively brief set of constraints is enough to define a legal measurement schedule.
However, most legal measurement schedules perform quite poorly -- certainly worse than the original LUCI algorithm.
In Section~\ref{sec:proxy} we consider how to capture the features we want a measurement schedule to exhibit using this model.

\subsection{Two- and Three-Round Measurement Schedules}\label{sec:shorter_schedules}
Although not sufficient to show an improvement in LER, it is noteworthy that the constraints in Section~\ref{sec:ilp-model} are enough to determine whether shorter measurement schedules exist.
One of the shortcomings of the LUCI framework is that, as originally formulated, it only guarantees that each mid-cycle operator will be measured once every four rounds of syndrome extraction.
This means that, in the worst case, it may require $4d$ rounds of syndrome extraction to suppress time-like logical errors to the same extent as space-like logical errors are suppressed.
By reducing the number of boards below the default $4$, i.e. bounding $t < 4$, we can relatively easily check whether the constraints in this section are satisfiable for a $2$- or $3$-round measurement schedule.
Although we do not perform detailed stability experiments to quantify the impact on time-like distance, we do note that 
$95/100(76/100)$ configurations sampled from $1\%(3\%)$ dropout on the $d = 11$ grid admit a $3$-round measurement schedule, reducing the required rounds to a worst case of $3d$.

It is important to recognize when shorter measurement schedules imply lower LER.
It is possible to argue that, presented with a three-round schedule and a four-round schedule, one should compare LER for $d$ blocks of three or four rounds respectively~\cite{wolanski2025automatedcompilationincludingdropouts}.
This may be justified by the fact that a three-round schedule performs the same logical operation as the four-round schedule (i.e. error correction).
When limited by time-like errors, e.g. during move operations or lattice surgery, this is a well-justified comparison, since in this situation one performs a fixed number of rounds in order to ensure time-like errors can not go undetected.
In the case of memory this comparison is not as natural.
In this scenario, one wishes to preserve logical information for a fixed period of time (e.g. idling a logical qubit while waiting for other qubits that need to interact with it to become available), not a fixed number of rounds of error correction.
The performance of preserving logical information for a fixed period of time is better captured by comparing the LER of two circuits with the same number of layers of physical operations, not two circuits with the same number of blocks of syndrome extraction.

In Section~\ref{sec:ilp-data} we briefly examine the LER performance of shorter measurement schedules, while in Appendix~\ref{appendix:3_round} we reason about the performance we should expect.

\subsection{A Linear Proxy for Logical Error Rate}\label{sec:proxy}
As foreshadowed earlier in this section our strategy for reducing LER is broadly to reduce detector volume, without creating detectors of unbalanced size.
To achieve this we rely upon five terms, which we spend the remainder of this section describing and justifying.

\begin{figure}
    \includegraphics[width=\linewidth]{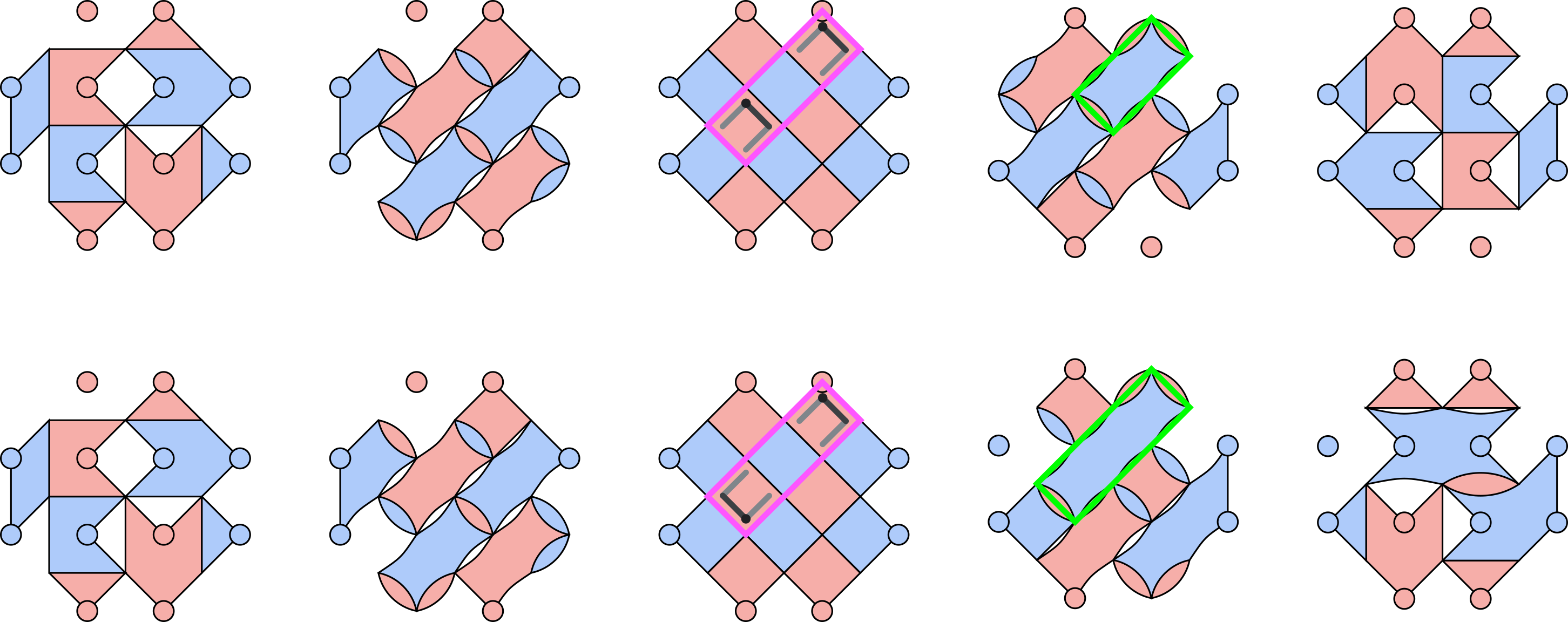}
    \caption{\label{fig:stretch}We compare the detectors for two plays which differ only in the orientation of one diagonal, with all but the two relevant shapes (fuchsia box) omitted in the mid-cycle state. We see that by opening two shapes towards each other, one detector becomes much larger (green box).}
\end{figure}

\begin{itemize}
    \item \textbf{Skip Twice} ($s_2$) counts the number of operators $i$ satisfying
    \[
        \exists_t : \lnot (f_{t, i} \lor f_{t + 1, i}).
    \]
    \item \textbf{Skip Thrice} ($s_3$) counts the number of operators $i$ satisfying
    \[
        \exists_t : \lnot (f_{t, i} \lor f_{t + 1, i} \lor f_{t + 2, i}).
    \]
\end{itemize}
To define the next term we must first formalize what it means for a shape to `stretch' a neighboring (super)stabilizer. 
We define $\text{stretches}_i(v_{t, j, p})$ to be \textsc{true} if using the shape $p$ to measure operator $j$ increases the spatial extent on which (super)stabilizer $i$ is sensitive to errors (and \textsc{false} otherwise).
This occurs when $p$ contains at least one \textsc{cnot} which is partially, but not completely, supported on the qubit support of stabilizer $i$, and the \textsc{cnot}s are the correct orientation to carry errors from neighboring qubits to the support of the stabilizer in question.
An example is given in Figure~\ref{fig:stretch}, in which we can see the differences in the spatial extent of a detector caused by the shapes used to measure neighboring operators.
Using this idea, we penalize any (super)stabilizer being stretched by more than one neighboring shape in a given round -- we do not penalize the case of exactly one since, broadly speaking, this is an inevitable consequence of measuring more than one neighboring operator in the same round, a behavior we wish to encourage.
\begin{itemize}
    \item \textbf{Alignment} ($a$) is defined as \[
        \sum_t \sum_i s_{t, i}
    \]
    where $s_i$ is constrained by \begin{align*}
        s_i &\geq 0\\
        s_i &\geq -1 + \sum_{x \in S_{t, i}} x
    \end{align*} and $S_{t, i} = \{v_{t, j, p} | \text{stretches}_i(v_{t, j, p})\}$ is the set of all shapes in round $t$ which increase the qubit support of (super)stabilizer operator $i$.
    Note that we are summing over boolean variables, but treating them simply as integers in $\{0, 1\}$ (not summing mod-2).
\end{itemize}
Each of these terms encodes one way in which stabilizers can be increased in size, or become unbalanced in time-like extent.

For longer measurement schedules (i.e. those with more than $4$ rounds) the natural extension to this set of terms would be to include $s_i, i > 3$.
It should also be observed that as written (and implemented), any contribution to the skip thrice term $s_3$ will also contribute to the skip twice term $s_2$.

Finally, we have seen empirically that the three-coupler surface code performs notably worse than the four-coupler surface code, but do not have as complete an explanation in terms of detector volume.
Rather, we note that the three-coupler measurement circuit directly allows for more minimum-weight paths through the graph.
We penalize these configurations using the following term:
\begin{itemize}
    \item \textbf{Basis Changes} ($b$) is defined as 
    \[\sum_{t} \sum_q (1 - J_{q, t}),\]
    where $J_{q, t}$ is the indicator variable encoding whether the basis that qubit $q$ is measured in is the same in rounds $t$ and $t + 1$ (defined to be zero if $q$ is not measured in both rounds).
\end{itemize}
For the three-coupler circuit, $J_{q, t}$ is always $0$, while for the four-coupler circuit $J_{q, t}$ is always $1$, allowing us to encode how similar a schedule is to each possibility.

We balance these terms encoding ways in which detectors can be made larger against a lower bound on the total number of deterministic measurements made.
All else being equal, we should expect that making a deterministic measurement should give us more information about any errors that have occurred, reducing LER.

\begin{figure}
    \centering
    \includegraphics[width=0.8\linewidth]{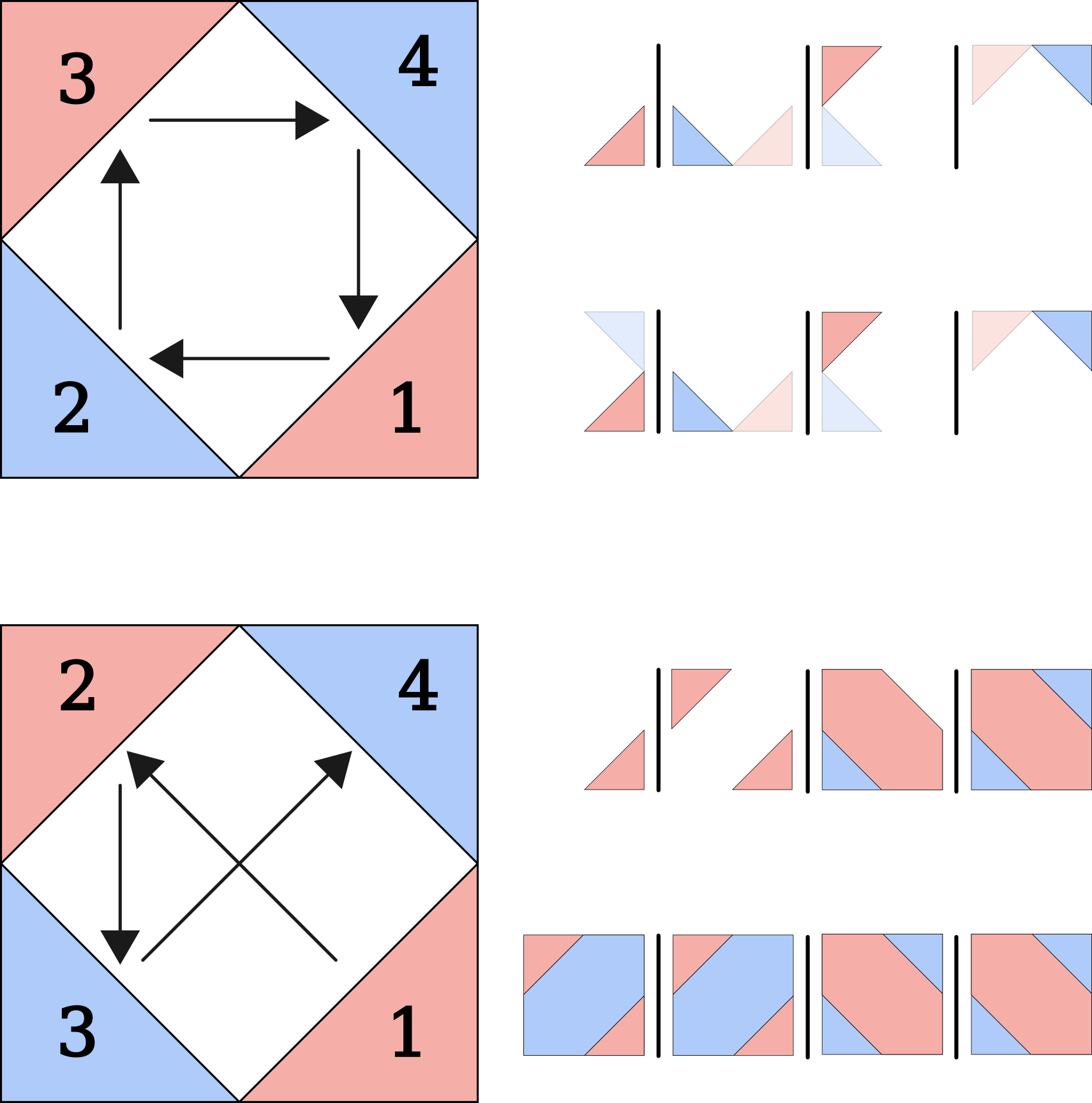}
    \caption{\label{fig:useless_measurements}Top left: An alternating $X/Z$ schedule fails to infer superstabilizer eigenvalues, as tracking the ISG over two rounds of measurement (top right) shows that each measurement removes the opposing gauge operator before it is measured (shown as a fading out of the removed gauge operator). 
    Bottom left: Grouping measurements by basis (e.g., all $X$ then all $Z$) resolves this issue, ensuring that half of the measurements are of operators currently present in the ISG.}
\end{figure}

\begin{itemize}
    \item \textbf{Deterministic Measurements} ($m$) is defined as 
    \[
        \sum_{t}\sum_{i \in S} f_{t, i} + \sum_{t}\sum_{i \in G}(f_{t - 1, i} \land f_{t, i}),
    \]
    where $S$ is the set of indices of stabilizer operators and $G$ is the set of indices of gauge operators.
\end{itemize}
This definition counts a measurement as contributing to $m$ if it is the measurement of a stabilizer, or of a gauge operator that is measured in the previous round; we do not count measurements directly because of the possibility of a measurement schedule similar to the one displayed in Figure~\ref{fig:useless_measurements}, in which four gauge operators are measured but no information on the eigenvalues of the superstabilizers they define is learned.
We wish to avoid this scenario, since the measurements are useless, they constrain neighboring measurements, and implementing the shapes used to measure them adds noise to the system.
Since, by Equation~\ref{eqn:measure_once}, we are guaranteed that each gauge operator is measured in some round $t$, the measurement of the same gauge operator in round $t + 1$ or $t - 1$ will increase $m$.
It is only a measurement in round $t + 2$ which may be both deterministic and not counted by $m$.
In principle, better determination of when the measurement of a gauge operator is deterministic could improve the numerical results presented in Section~\ref{sec:ilp-data}.

With these terms in hand, then, we simply minimize the objective function
\begin{equation}
        -m + \alpha\times s_2 + \beta\times s_3 + \gamma\times a + \delta\times b,
\end{equation}
where we have fixed the coefficient of $m$ to $-1$, effectively normalizing $s_2, s_3, a, b$.

%TODO: adjust if we change the parameters
We have now reduced our optimization problem to finding appropriate values of $\alpha, \beta, \gamma, \delta$. 
In numerical experiments, the results of which are summarized in Section~\ref{sec:ilp-data}, we used $\alpha, \beta, \gamma, \delta = 6, 5, 12, 2$.
We justify the relation between $\alpha, \beta, \gamma$ by assuming that the impact on LER of any one detector is roughly linear in its volume and that most operators we measure are weight-four.
Therefore, the effect of skipping a measurement is to add $4$ locations to the detector it would otherwise have finished, while the effect of stretching an operator is to add $2$ locations -- this implies that $\alpha = 2\gamma$. 
We then assume that skipping a measurement three times is approximately $3/2$ as harmful as skipping it twice, but reduce the coefficient slightly to account for the fact that most schedules in which a measurement is skipping three times in a row do not also stretch the operator in all skipped rounds.
We have less justification for the value of $\delta$ chosen, except the intuition that it is the smallest effect out of the ones considered.

Intuitively, we can understand this objective function as encoding tradeoffs between space-like and time-like dimensions.
The alignment term captures space-like extent, ensuring that no detector gets too large in a single time-slice while the skip terms capture time-like behavior, ensuring that detectors that reach too far in time are penalized.
The measurement term is included because we assume that adding any deterministic measurement, all else being equal, is generally positive. 

We solve this ILP model using CP-SAT from Google's OR-tools~\cite{cpsatlp}.
A critical consideration for this approach is computational feasibility.
For the $d = 11$ surface code instances studied, involving approximately $13,000$ variables and $60,000$ constraints (including auxiliary variables and constraints), provably finding an optimal solution can be impractical.
We have observed, however, that finding the optimal solution is relatively quick (on the order of $5$ minutes on a standard workstation), while it is \emph{proving} that no better solution can be found that requires the majority of the time.
Therefore, for the numerical experiments in Section~\ref{sec:ilp-data} we have limited the time provided to the solver to $5$ minutes.
We discuss convergence rate and improvements to LER due to longer time limits in more detail in Section~\ref{sec:ilp-data}.

It also greatly improves solution quality under time-limits to provide a high-quality feasible solution as a set of hints to the solver -- we use the schedule provided by the algorithm in Section~\ref{sec:background} to seed the solver, which also has the effect of ensuring that the final solution will be of no worse quality (relative to the objective function we have set) than the original solution.
While computationally intensive, this is a one-time optimization practical for designing a high-performance circuit.

\section{Numerical Impact of Circuit Optimization}\label{sec:ilp-data}
\begin{figure}
    \centering
    \begin{subfigure}[b]{\linewidth}
        \centering
        \includegraphics[width=\linewidth]{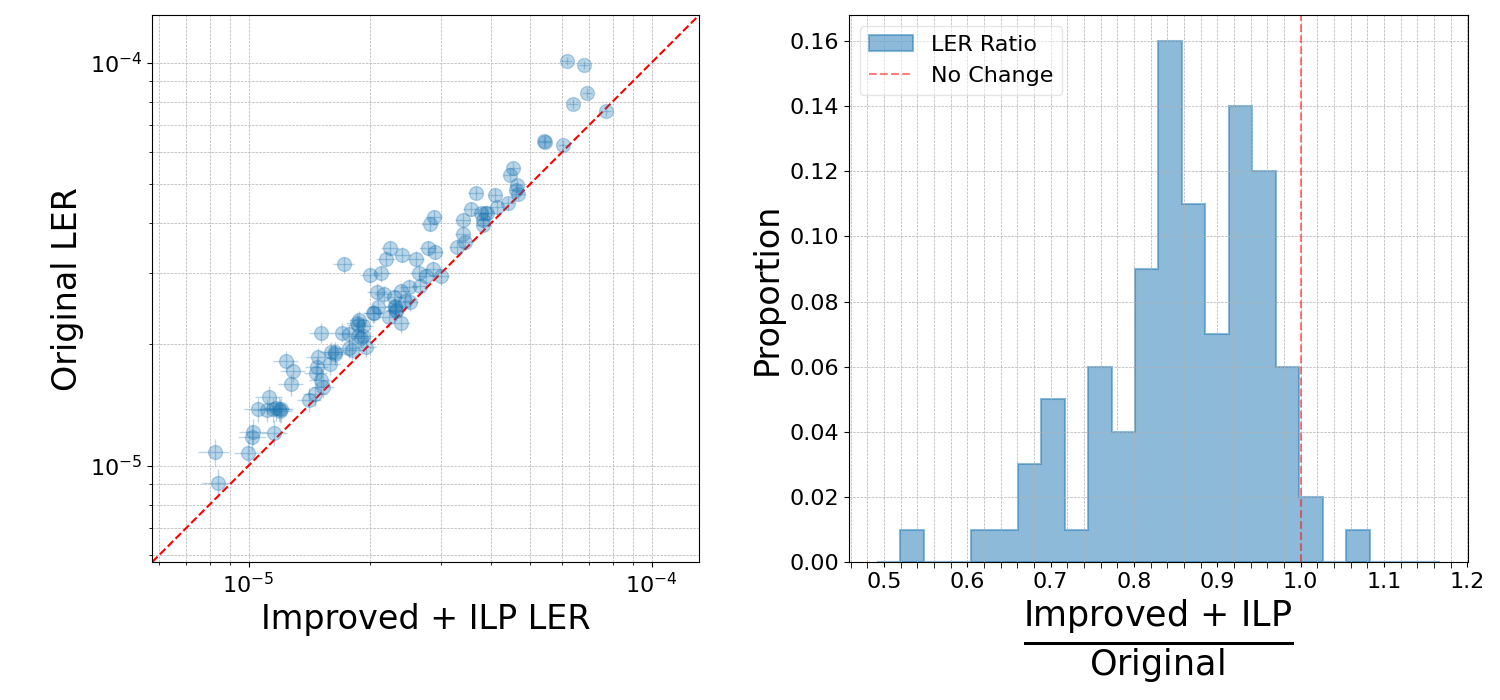} 
        \caption{LER comparison at $1\%$ dropout.}
        \label{10a}
    \end{subfigure}
    \begin{subfigure}[b]{\linewidth}
        \centering
        \includegraphics[width=\linewidth]{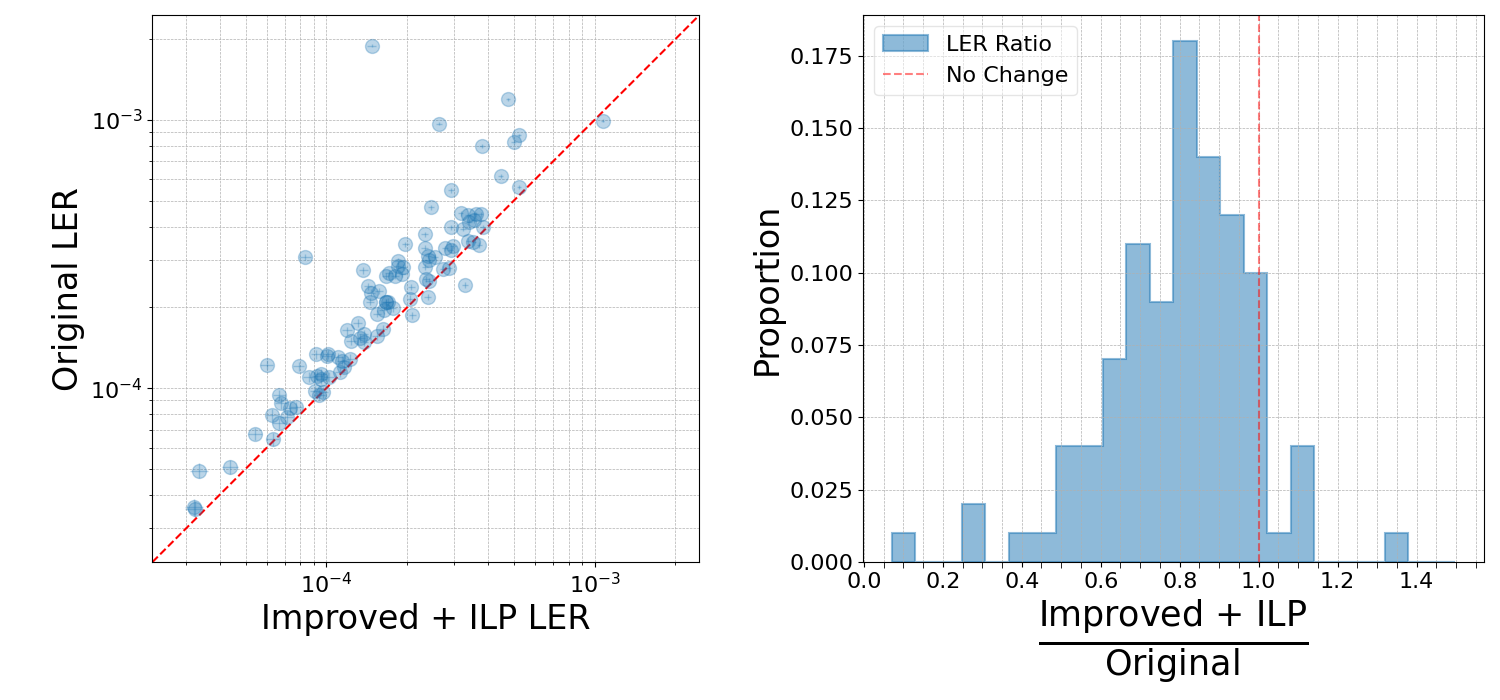}
        \caption{LER comparison at $3\%$ dropout.}
        \label{10b}
    \end{subfigure}
    \begin{subfigure}[b]{0.48\linewidth}
        \centering
        \includegraphics[width=\linewidth]{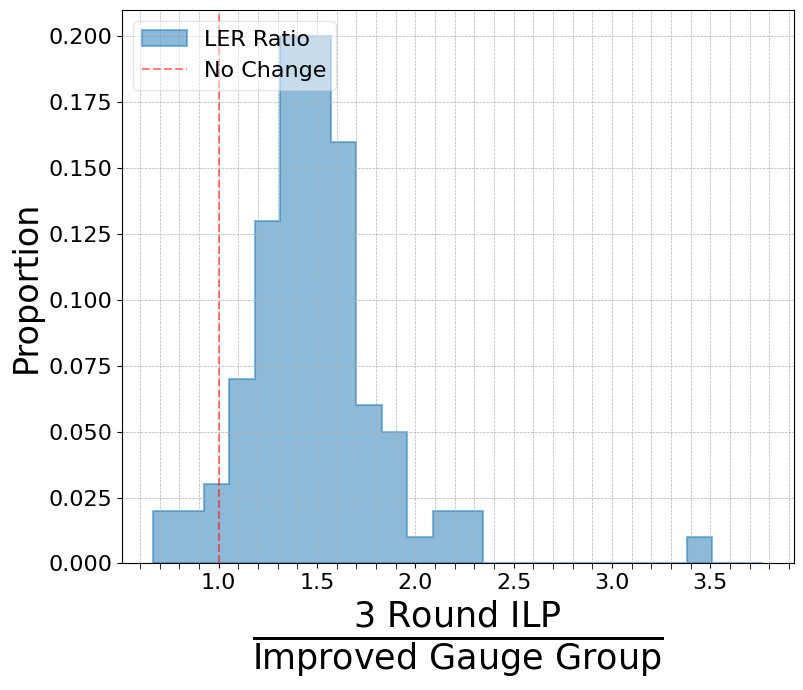} 
        \caption{LER comparison at $1\%$ dropout for a three-round schedule.}
        \label{10c}
    \end{subfigure}
    \begin{subfigure}[b]{0.48\linewidth}
        \centering
        \includegraphics[width=\linewidth]{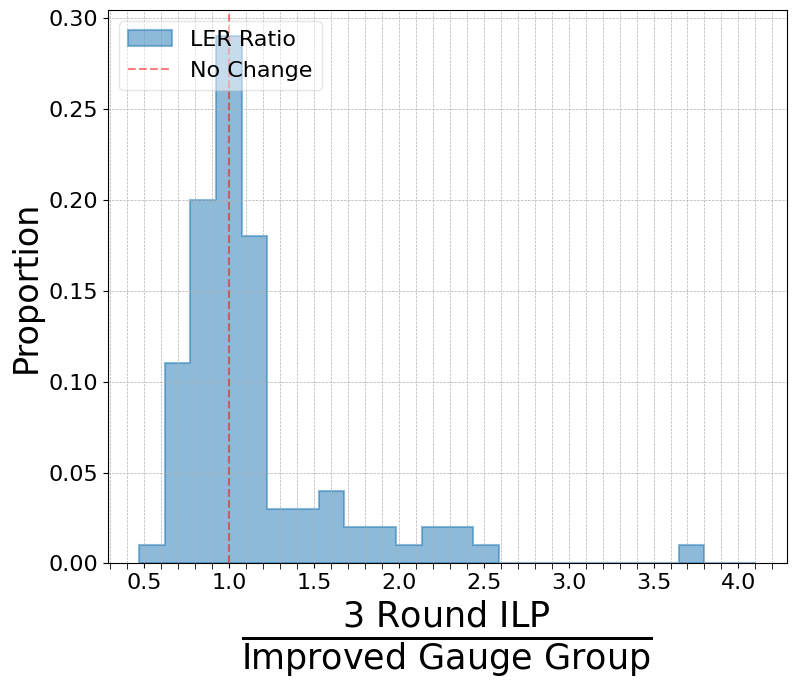}
        \caption{LER comparison at $3\%$ dropout for a three-round schedule.}
        \label{10d}
    \end{subfigure}
    \caption{\label{fig:ilp-data}
    Comparison of LER for $d = 11$ using the SI1000 noise model~\cite{gidney2021honeycomb} at $0.1\%$ physical noise rate with and without optimizations of both the operators and the schedule we use to measure them.
    On the left of Figures~\ref{10a},~\ref{10b} we directly compare LER for each grid with $95\%$ confidence intervals marked, while on the right we plot the distribution of LER ratios. 
    We see that we can enjoy substantial decreases to LER by optimizing the play using the improvement in Section~\ref{sec:wt-one-gauges} and the circuit according to the objective function proposed in Section~\ref{sec:proxy}.
    In Figures~\ref{10c} and~\ref{10d}, we see that three-round schedules generally perform worse than their unoptimized four-round counterparts.
    }
\end{figure}
Before we examine the features of the optimized schedules, we first present the overall results.
We see in Figure~\ref{fig:ilp-data} that optimizing according to the objective function from Section~\ref{sec:proxy} using the model proposed in Section~\ref{sec:ilp-model} yields consistent improvement in LER in both the $1\%$ and $3\%$ dropout regimes.
The reduction in geometric mean (relative to the simulations from Section~\ref{sec:wt-one-gauges}) is $6.8\%(10.3)\%$, for the $1\%(3\%)$ dropout case, demonstrating that substantial increases to logical lifetime can be had without requiring any extra hardware capabilities (or even changing which stabilizers are measured).

Notably, we observe an \emph{increase} in LER when using the same ILP model but restricting to three-round solutions.
Such solutions produce an increase in LER of $45\%(8.1\%)$ at $1\%(3\%)$ dropout, rather than the decreases observed when using four-round solutions.
We examine some of the reasons why we observe this behavior in Appendix~\ref{appendix:3_round}.
The significant difference in LER increase between $1\%$ and $3\%$ dropout is because the $3\%$ four-round baseline is already relatively constrained, with approximately $30\%$ of operators only able to be measured once. 
Switching to three rounds increases this to approximately $60\%$, a less dramatic relative shift than in the $1\%$ case, where single-measurement frequency jumps from $10\%$ to $50\%$. 

Despite the increased LER, such schedules may still be useful in the case that one is limited by time-like distance, e.g. after lattice surgery.
Depending on the exact impact of time-like errors, there are scenarios in which reducing the required time to perform a lattice surgery operation by $25\%$ could reduce the total LER of the circuit, despite the fact that the specific logical qubits in question suffer from higher space-like LER.

Numerical experiments in this section use the same setup as those in Section~\ref{sec:wt-one-gauges}, i.e. taking $30$ million shots for $1\%$ and $10$ million for $3\%$, with circuits consisting of $4d$ rounds of syndrome extraction.

In the no-dropout case minimizing the objective function provided obtains one of the four symmetrical variations of the canonical surface code syndrome extraction circuit.
The fact that the reduction in LER is monotonically increasing in dropout rate in the $0\%, 1\%, 3\%$ cases suggests that optimized circuits are even more important in the very-high-dropout regime (a supposition which cursory examination of the $5\%$ dropout case also supports).
\subsection{Analysis of Optimized Schedules}
Having demonstrated the quantitative utility of optimizing error correction circuits using the LUCI framework, we now examine the qualitative behavior of our approach.
First, we examine whether our optimization actually succeeds in reducing detector volumes and the proportion of very high volume detectors, as these were the figures of merit we argued that the objective function in Section~\ref{sec:proxy} would capture.

\begin{figure}
        \centering
    \begin{subfigure}[b]{0.49\linewidth}
        \centering
        \includegraphics[width=\linewidth,  clip]{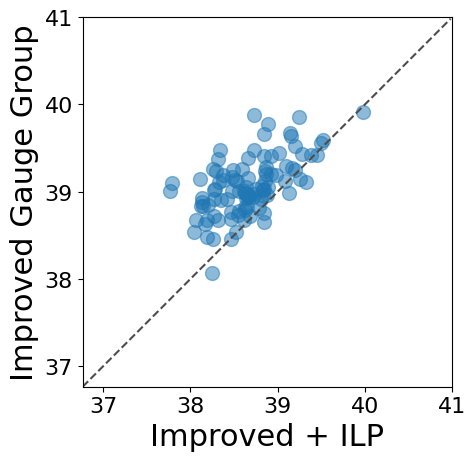} 
        \caption{Average detector volume for each of 100 grids sampled from $1\%$ dropout.}
    \end{subfigure}
    \begin{subfigure}[b]{0.49\linewidth}
        \centering
        \includegraphics[width=\linewidth, , clip]{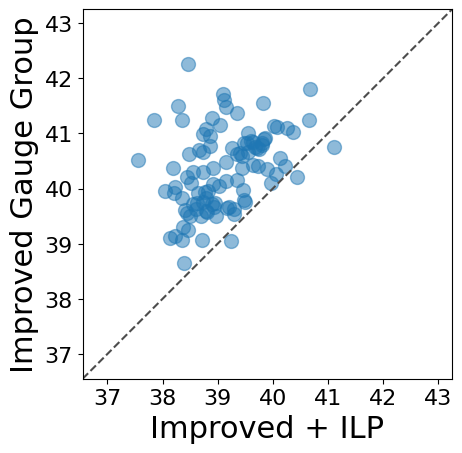}
        \caption{Average detector volume for each of 100 grids sampled from $3\%$ dropout.}
    \end{subfigure}\\
    \begin{subfigure}[b]{\linewidth}
        \centering
        \includegraphics[width=\linewidth]{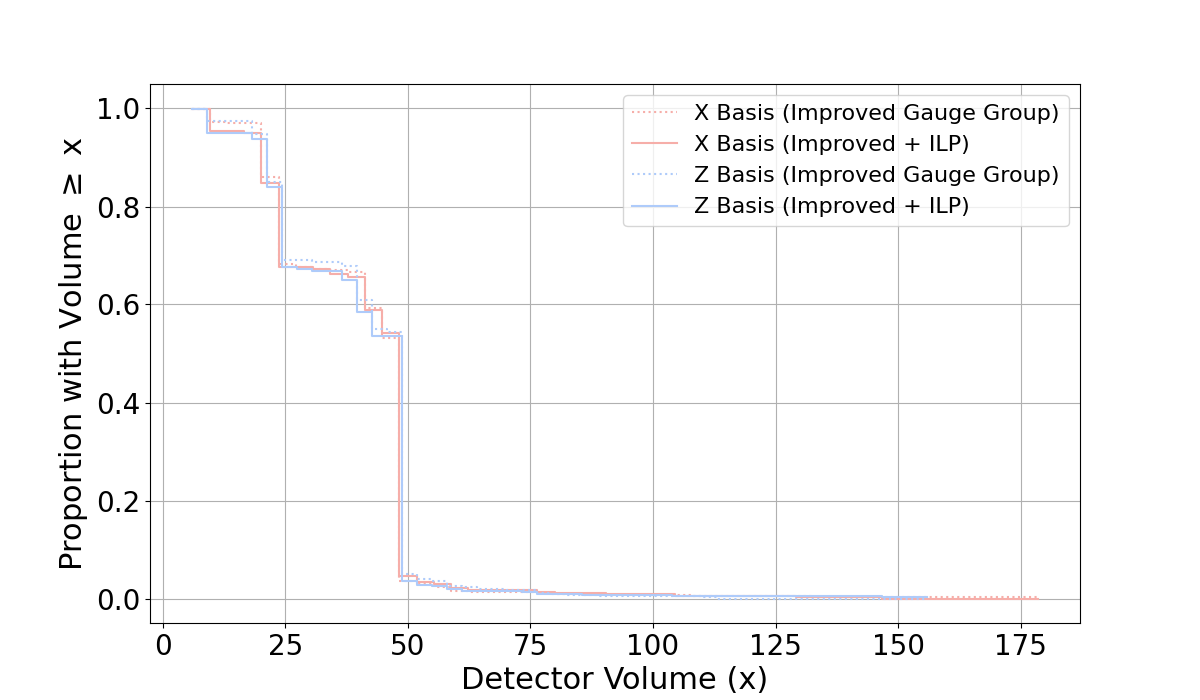}
        \caption{Reverse cumulative distribution of detector volumes for the default schedule (dotted) versus the optimized schedule (solid) for one representative dropout configuration.}
    \end{subfigure}
    \caption{We see that optimizing our measurement schedule using the objective function and model from Section~\ref{sec:ilp-model} does indeed produce smaller detectors.
    Average detector volume is reduced by $1.01\%(3.04\%)$ at $1\%(3\%)$ dropout.}
    \label{fig:med_det_vol}
\end{figure}

In Figure~\ref{fig:med_det_vol} we compare average detector volumes (where the volume of a detector is the total number of space-time locations it detects an error at) for the two schedules (optimized versus default with optimizations to the gauge group) produced for each grid as well as a representative example of the cumulative distribution of detector volumes, and observe that indeed our optimization has produced smaller, higher information, detectors.

\begin{figure}
        \centering
    \begin{subfigure}[b]{0.48\linewidth}
        \centering
        \includegraphics[width=\linewidth]{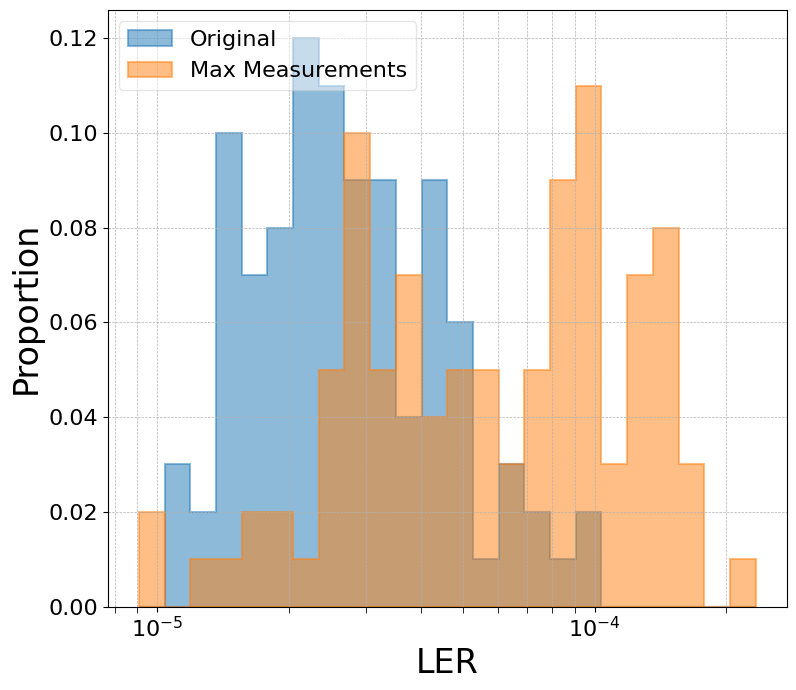} 
        \caption{LER comparison at $1\%$ dropout.}
    \end{subfigure}
    \begin{subfigure}[b]{0.48\linewidth}
        \centering
        \includegraphics[width=\linewidth]{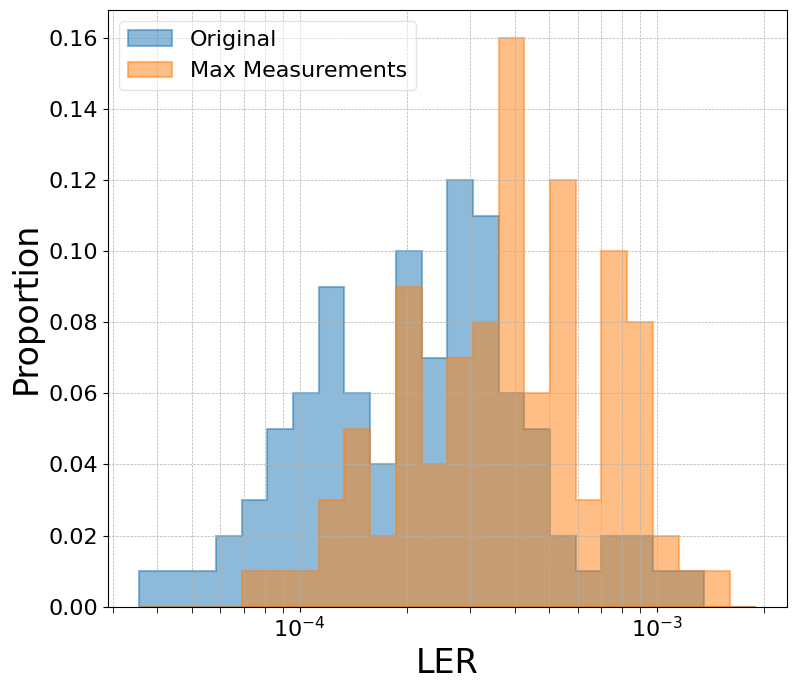}
        \caption{LER comparison at $3\%$ dropout.}
    \end{subfigure}
    \caption{\label{fig:max_vs_orig}
    Comparison of LER for $d = 11$ using the SI1000~\cite{gidney2021honeycomb} noise model at $0.1\%$ physical noise rate for unoptimized LUCI vs an ILP obtained schedule using the number of measurements made as its only metric.}
\end{figure}

We also recognize that it may be unintuitive that a na\"ive objective function based solely on maximizing measurements is insufficient.
In Figure~\ref{fig:max_vs_orig} we present data showing that not only is an objective function which solely maximizes measurements outperformed by the one we propose, it actually increases LER relative to vanilla LUCI (even when benefiting from the optimizations in Section~\ref{sec:wt-one-gauges}).
These results are similar to those of Appendix~\ref{appendix:acid}.

We also analyze how increasing time limits contribute to logical error rate.
\begin{figure}
    \includegraphics[width=\linewidth]{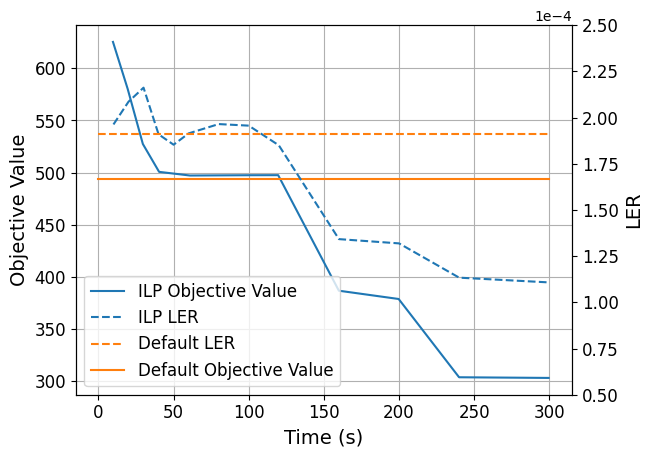}
    \caption{\label{fig:convergence}Time versus objective value and LER.
    We see that the objective value tends to be a good predictor of LER, but the relationship is not completely monotonic.
    For this experiment we do not provide hints to the solver, meaning that the objective value of the optimized schedule is not upper bounded by the default objective value at all times.}
\end{figure}
%TODO: adjust if we change the data
A representative example is shown in Figure~\ref{fig:convergence}.
There are several notable features of this graph.
We observe that in general increasing the time provided to the solver, and hence decreasing the objective value, does decrease logical error rate.
This relationship is not completely monotonic though; in particular, between seconds 50 and 120, we observe a LER that increases before decreasing again, while the objective value remains almost invariant.
This demonstrates that there can be multiple schedules with the same objective value but with different logical error rates, even for a fixed circuit distance/dropout configuration.
This is somewhat to be expected since LER is highly complex, and capturing its entire behavior with a simple linear objective function is a daunting task.
A deeper analysis of LER versus objective value could yield valuable insight as to what terms the objective value may be missing.

\section{Discussion and Future Work}\label{sec:conclusion}
We have demonstrated that, beyond serving simply as a single-point solution for handling fabrication defects, LUCI can be used as a powerful and expressive intermediate representation, which enables practical optimization both of the gauge group and of the physical circuit used to implement the corresponding error correction procedure.

Our results on the $d=11$ surface code highlight two distinct regimes of improvement. 
First, the inclusion of weight-one gauge operators provides a robust baseline improvement, particularly effective in high-dropout regimes where avalanche-style effects can easily occur, while the removal of effectively unused qubits after circuit design provides a noticeable improvement to logical error rate. 
Second, the ILP-based circuit optimization yields a significant further reduction in logical error rate. 
Crucially, this second improvement is achieved without changing the code distance or the stabilizers measured; it is purely a result of scheduling operations to maximize detector strengths.
The fact that our solver minimizes a proxy objective, based on detector volume and alignment, and yet achieves consistent LER reduction confirms that our physical intuition regarding ``detector stretching'' and ``measurement skipping'' maps correctly to logical performance.

%TODO: adjust if we change the parameters
A key feature of this approach is its robustness. 
The hyperparameters used in our objective function ($\alpha, \beta, \gamma, \delta$) were chosen based on first-principles physical arguments regarding space-time volume, rather than exhaustive hyperparameter tuning. 
This suggests that the method is not overfitting to a specific noise model or layout, but rather correcting fundamental inefficiencies in the heuristic schedule. 
However, this also points toward a promising avenue for future work: the systematic optimization of these weights. 
It is likely that a Bayesian approach could refine the coefficients of our objective function to yield even greater reductions in logical error rate.

The objective function could also be improved by including second-order (non-linear) terms.
Intuitively, two weak detectors in the support of the same minimum-weight representative of a logical operator are worse than weak detectors in the support of two distinct minimum weight representatives.
In this work, we have chosen not to include such terms due to the increased demand on solver time that linearizing quadratic penalties such as these represents, but in principle such an approach could yield better logical error rates. 

As implemented, this optimization framework applies only to the surface code (hex-grid or otherwise).
In principle, it is feasible to apply this framework to finding measurement schedules for other codes which well-behaved mid-out descriptions, such as bivariate bicycle codes~\cite{bravyi2024bb, shaw2025morphing}.
Indeed, such a strategy has been implemented recently in Ref.~\cite{wolanski2025automatedcompilationincludingdropouts}, but we expect that adapting the heuristics we have used to minimize and equalize detector volume would greatly improve LER.
Generalizing these heuristics from the surface code to higher-connectivity codes is a non-trivial open problem. 

While solving the ILP formulation is computationally intensive, using minutes of solver time for distance-11 grids, this cost is paid only once per device calibration cycle (or upon the identification of new defects). 
Given the static nature of fabrication defects, this one-time compilation cost is relatively small compared to the lifetime of the device.

Ultimately, this work establishes that as quantum hardware scales and fabrication defects become statistically certain, circuit optimization can be a significant part of bridging the gap to reliable quantum computation and good intermediate representations allow powerful classical optimization techniques to be applied.
\section{Acknowledgements}
We thank Adam Zalcman, Matt McEwen, Noah Shutty and Oscar Higgott for thoughtful discussion and advice.
\onecolumngrid
\clearpage
\twocolumngrid
\bibliographystyle{apsrev}
\bibliography{refs}
%\onecolumngrid
\clearpage
\appendix
\section{Additional Data}
In this section we include various supplementary graphs showing the impact of the two optimizations considered separately.
\begin{figure}[h!]
    \centering
    \begin{subfigure}[b]{\linewidth}
        \centering
        \includegraphics[width=\linewidth]{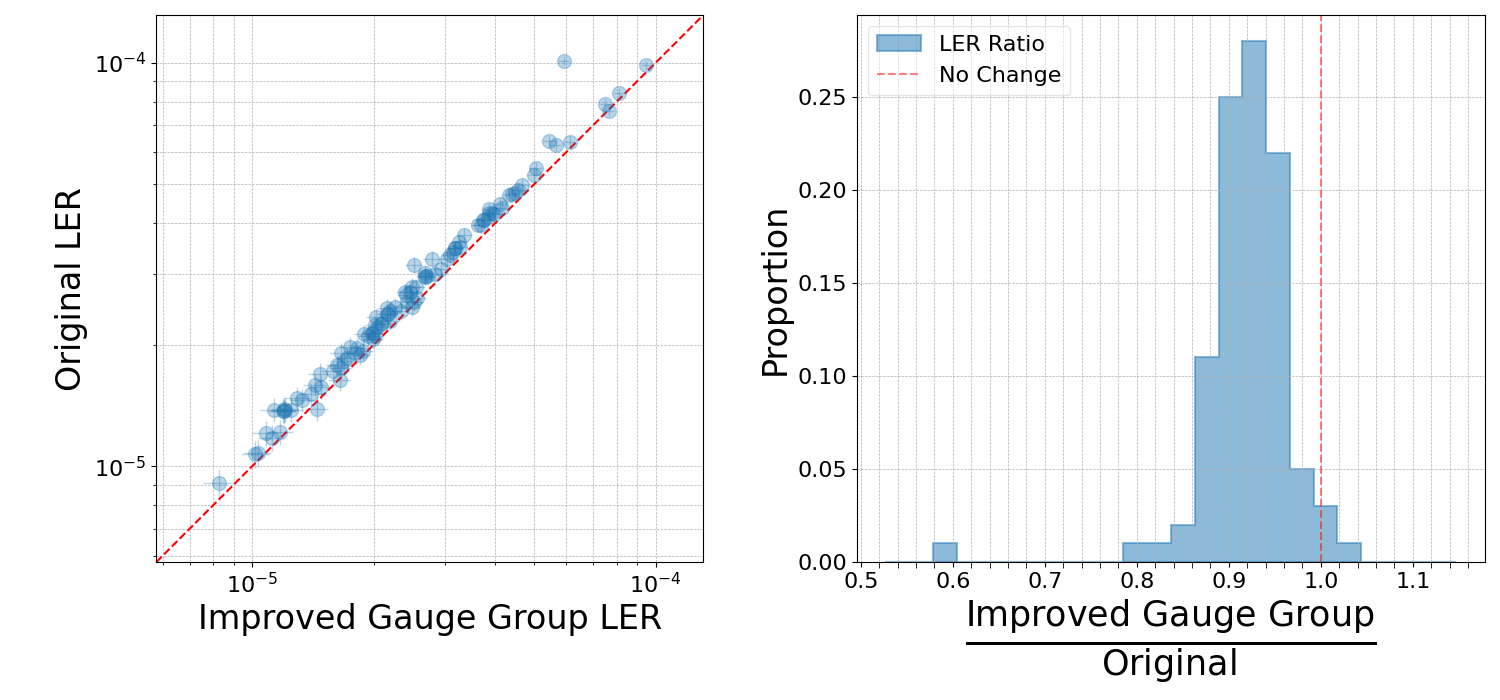} 
        \caption{LER comparison at $1\%$ dropout.}
    \end{subfigure}
    \begin{subfigure}[b]{\linewidth}
        \centering
        \includegraphics[width=\linewidth]{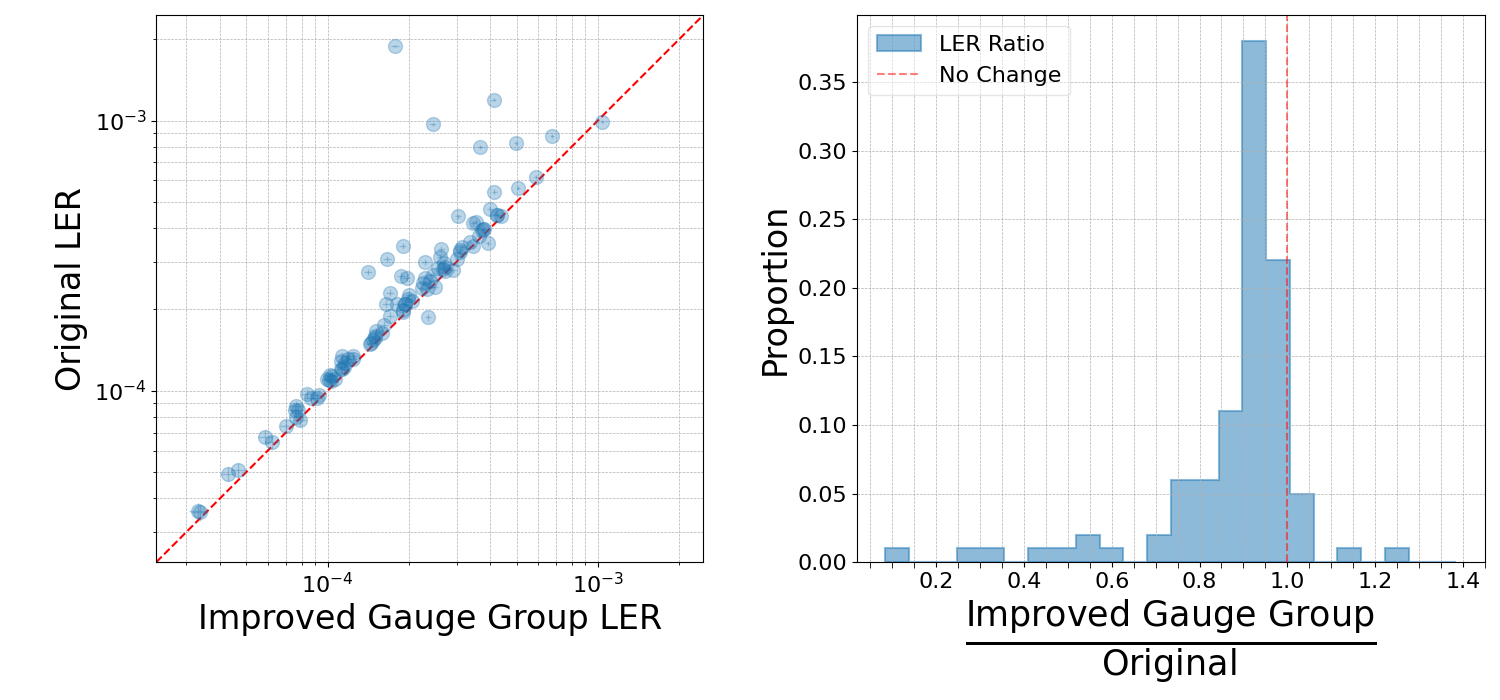}
        \caption{LER comparison at $3\%$ dropout.}
    \end{subfigure}
    \caption{
    Comparison of LER for $d = 11$ using the SI1000~\cite{gidney2021honeycomb} noise model at $0.1\%$ physical noise rate when including weight-one gauge operators and removing weight-one stabilizer operators where possible versus the original algorithm.
    On the left we directly compare LER for each grid with $95\%$ confidence intervals marked, while on the right we plot the distribution of ratios.
    We see that although the impact of improving the choice of gauge group is most extreme at higher dropout rates, it is beneficial in both of the regimes we consider.}
    \label{fig:wt-one-data}
\end{figure}
\begin{figure}
    \centering
    \begin{subfigure}[b]{\linewidth}
        \centering
        \includegraphics[width=\linewidth]{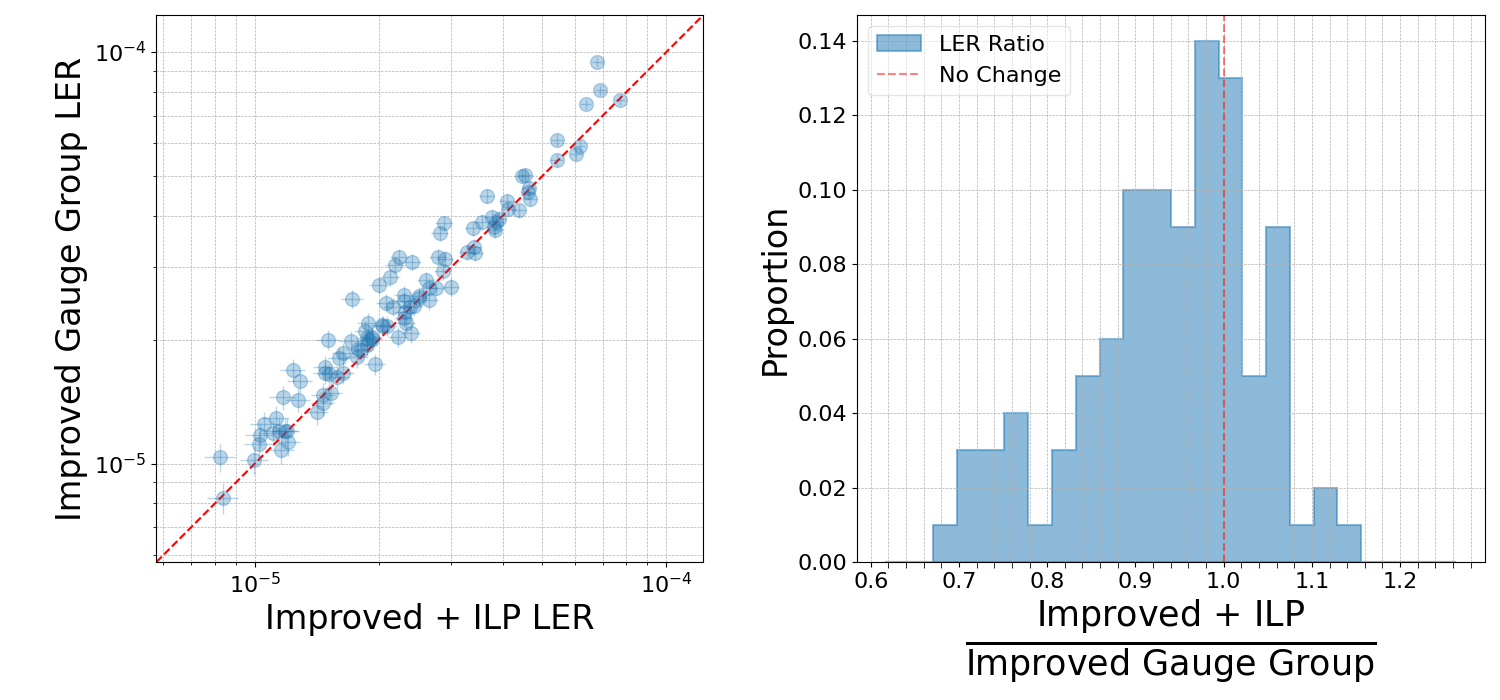} 
        \caption{LER comparison at $1\%$ dropout.}
    \end{subfigure}
    \begin{subfigure}[b]{\linewidth}
        \centering
        \includegraphics[width=\linewidth]{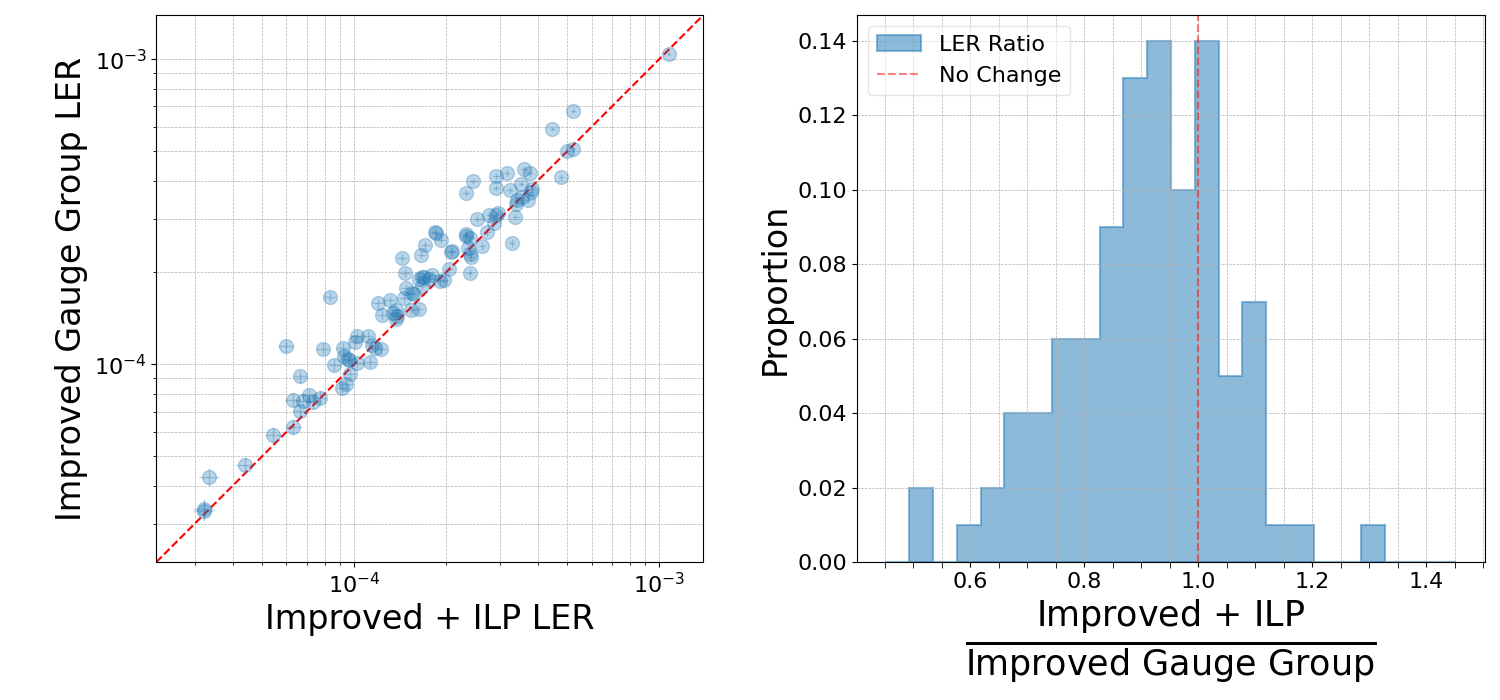}
        \caption{LER comparison at $3\%$ dropout.}
    \end{subfigure}
    \caption{\label{fig:ilp-data-appendix}
    Comparison of LER for $d = 11$ using the SI1000 noise model~\cite{gidney2021honeycomb} at $0.1\%$ physical noise rate with and without circuit optimization.
    On the left we directly compare LER for each grid with $95\%$ confidence intervals marked, while on the right we plot the distribution of LER ratios.    
    We see that we can enjoy substantial decreases to LER by optimizing the circuit according to the objective function proposed in Section~\ref{sec:proxy}, using the same dropout configurations as in Figure~\ref{fig:wt-one-data}.}
\end{figure}
\begin{figure*}[t]
    \begin{subfigure}{0.75\linewidth}
        \centering
        \includegraphics[width=\linewidth]{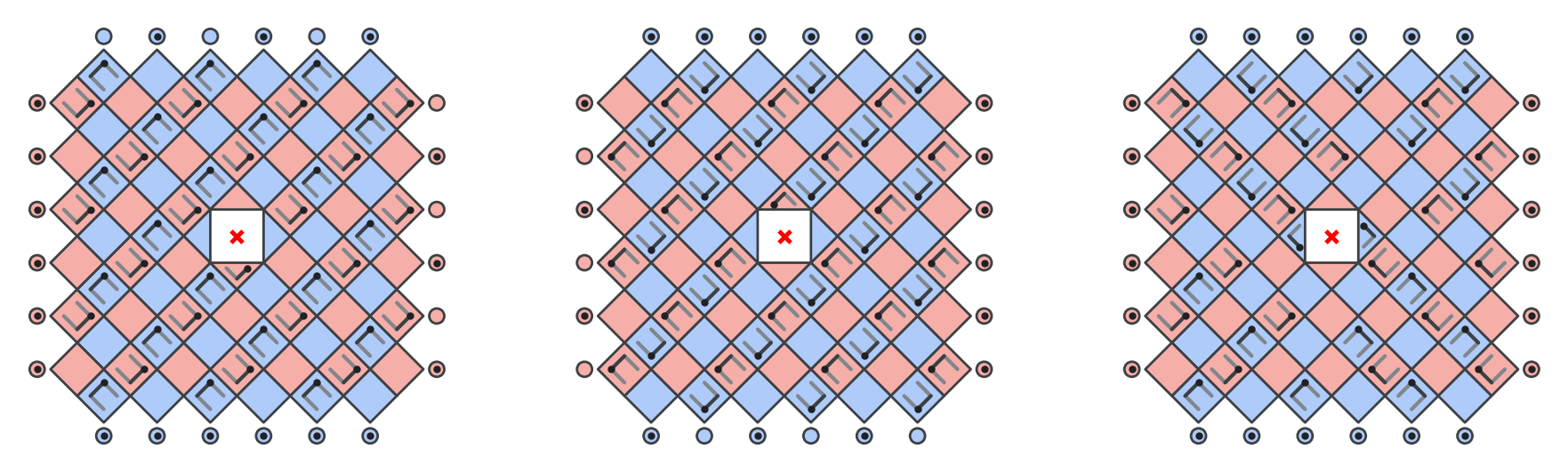} 
        \caption{A three round measurement schedule.}
    \end{subfigure}
    \begin{subfigure}{\linewidth}
        \centering
        \includegraphics[width=\linewidth]{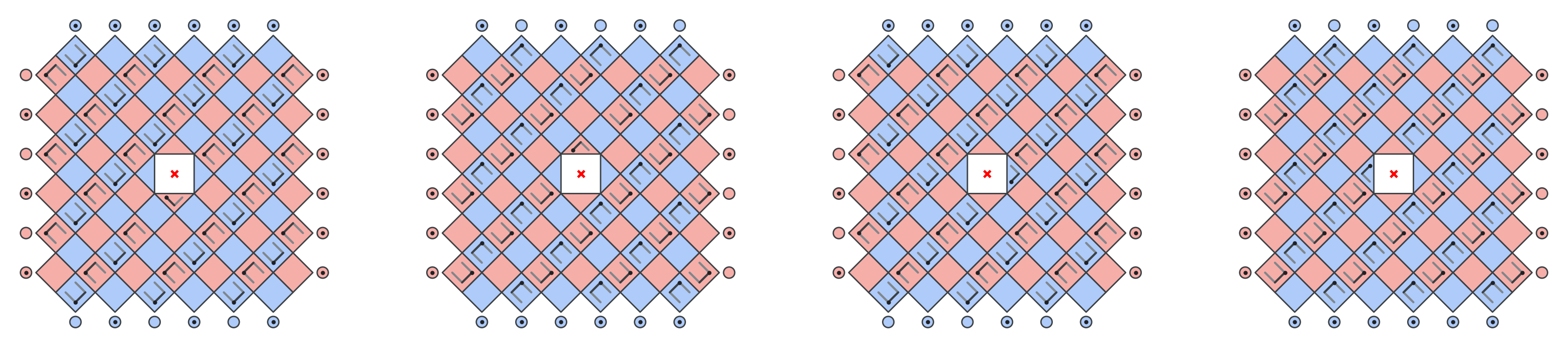}
        \caption{A four-round measurement schedule.}
    \end{subfigure}
    \caption{Comparison of three- and four-round measurement schedules for a simple dropout configuration.}
    \label{fig:windmill}
\centering
\end{figure*}
\section{Analysis of Three-Round Schedules}\label{appendix:3_round}
In Section~\ref{sec:ilp-data}, we observe that three-round schedules are generally worse than four-round schedules.
This fact may be surprising, since three-round schedules have a circuit which is $25\%$ shorter than those of four-round schedules, intuitively suggesting that their LER should also be approximately $25\%$ lower.

This apparent discrepancy can be explained by considering the measurement frequency of each operator.
While it is true that a three-round schedule guarantees that each operator is measured at least once every three rounds, as opposed to the once every four rounds guaranteed by a four-round schedule, a three-round schedule produces many more operators saturating this bound than the four-round case.

Intuitively, most of the operators in a four-round schedule are measured once every second round.
This fact can be seen by considering the algorithm outlined in Section~\ref{sec:background} and noting that isolated cases of dropout halve the measurement frequency of surrounding operators (from once every two rounds to once every four rounds), leaving the frequency of operators not adjacent untouched.
Quantitatively, each qubit is surrounded by 4 operators, while each coupler is surrounded by 6, meaning that at $1\%(3\%)$ dropout we should expect approximately $10\%(30\%)$ of the operators to only be measured once.

For a three-round schedule, we analyze it by fixing the first two rounds.
The third round then must measure all of the operators not measured in the first two rounds, as well as filling in as many extra operators as possible (roughly speaking).
Since rounds one and two are incompatible, the third round must (locally) pick between operators measured in round one and round two, meaning approximately $50\%$ of the operators are only measured once every three rounds.
This is on top of the reduced local measurement frequency implied by dropout.

In practice, we observe this manifesting as a `windmill' effect, where rounds one and two hold mostly diagonal orientations and round three holds a mostly antidiagonal orientation (or vice versa) as in Figure~\ref{fig:windmill}.
In this example, $50$ of the $84$ bulk operators are measured only once in the three-round schedule, while only $6$ are measured less frequently than once every other round in the four-round schedule.
Despite the fact that no detector in the three-round schedule has the same time-like extent as the longest detectors in the four-round schedule, most detectors are larger in the three-round schedule.
\section{Case study of Maximizing Measurements}\label{appendix:max_meas_bad}
\begin{figure*}
        \centering
        \begin{subfigure}{\linewidth}
            \centering
            \includegraphics[width=\linewidth]{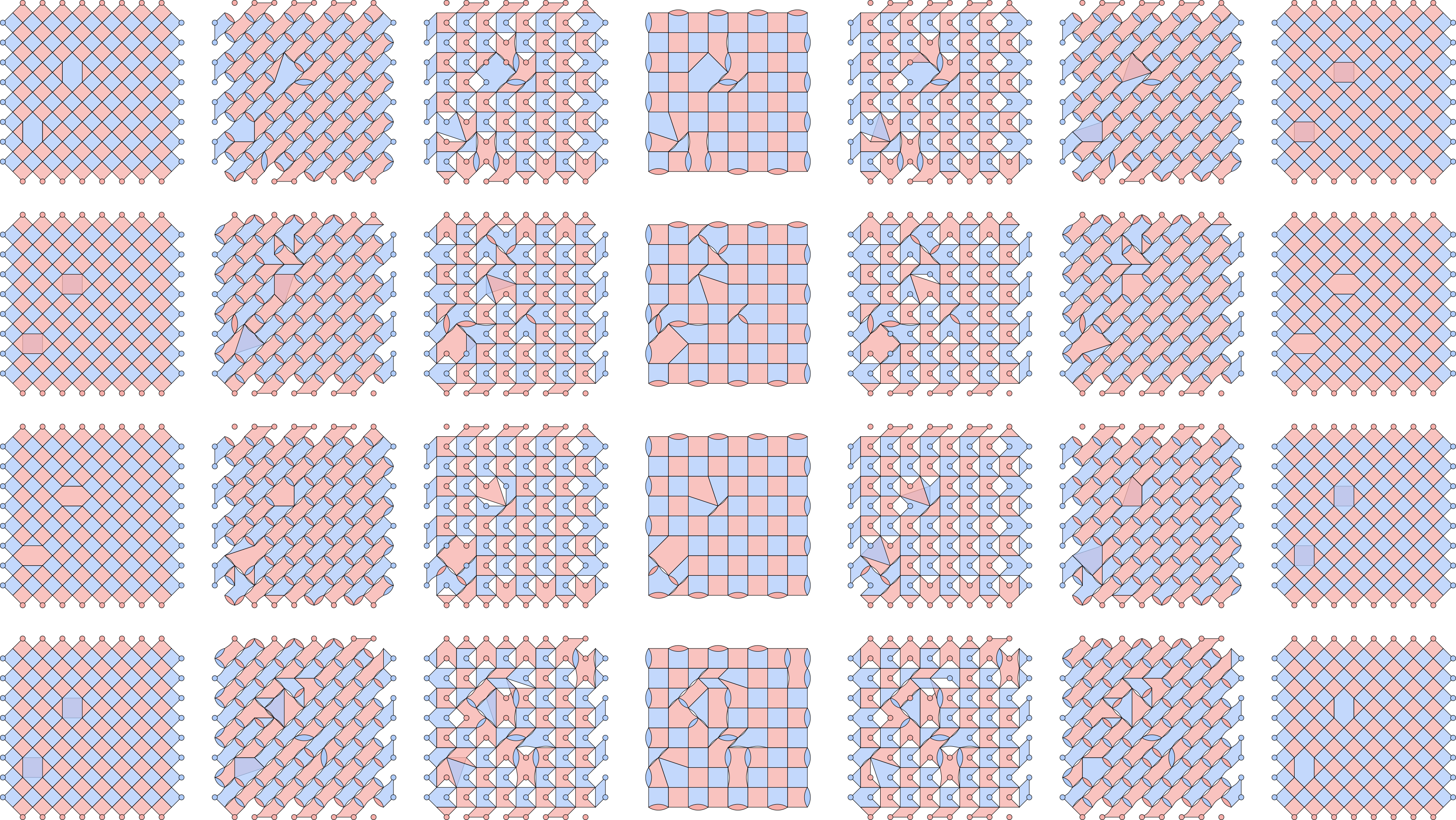} 
            \caption{Detector slices for Figure~\ref{fig:schedule_comparison}a.}
        \end{subfigure}
        \begin{subfigure}{\linewidth}
            \centering
            \includegraphics[width=\linewidth]{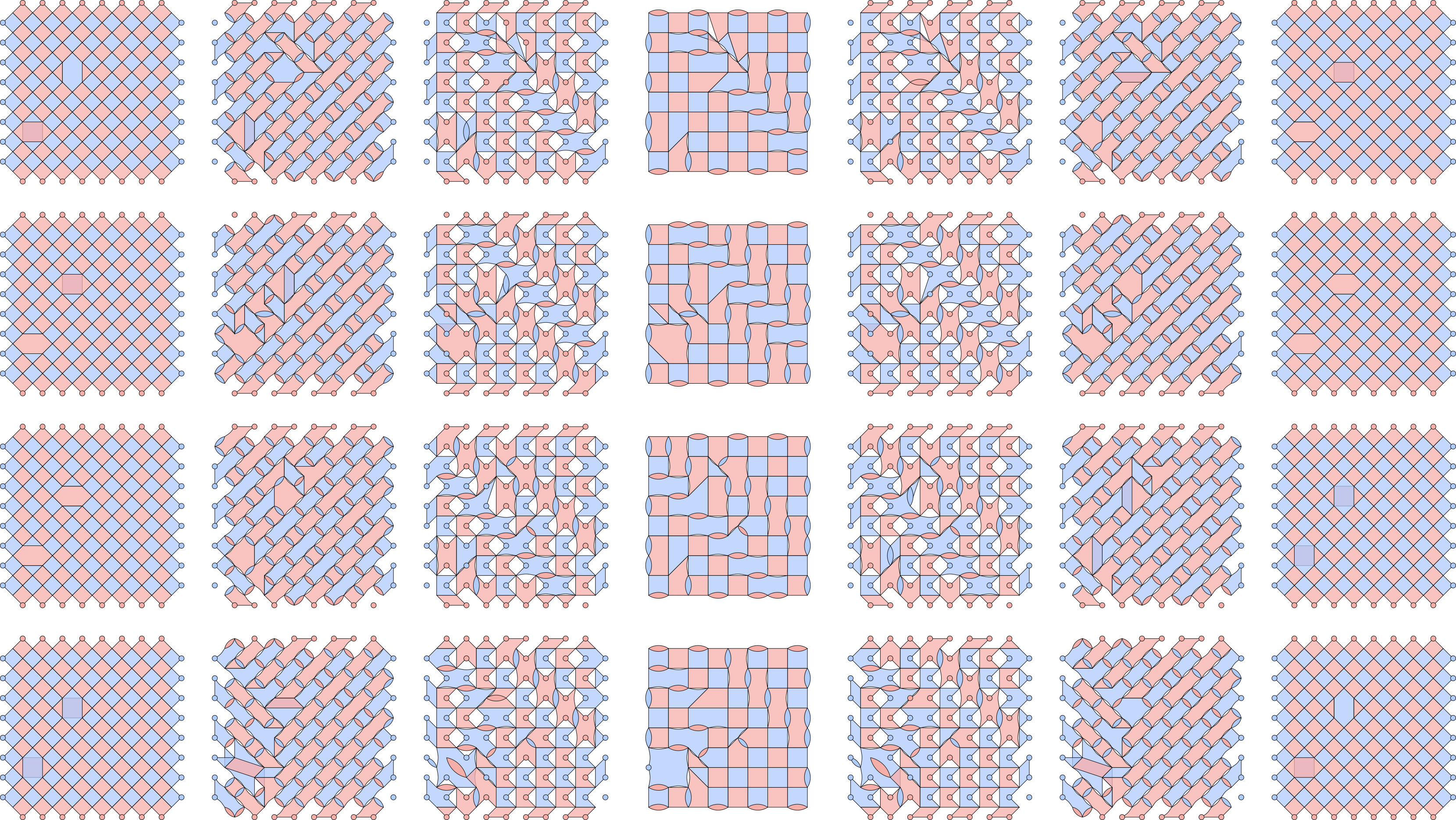}
            \caption{Detector slices for Figure~\ref{fig:schedule_comparison}c.}
        \end{subfigure}
        \caption{\label{fig:detslice_comparison}Comparison of detector slices for two of the candidate schedules in Figure~\ref{fig:schedule_comparison}.
        Each row corresponds to one board (i.e. mid-cycle to mid-cycle round).}
\end{figure*}

In Figure~\ref{fig:schedule_comparison} we presented two plays with significantly different LER.
Here we examine the structure of the circuits produced to help explain why one performs worse than the other.
In Figure~\ref{fig:detslice_comparison} we directly compare the detectors implied by the schedule in Figure~\ref{fig:schedule_comparison}(a) and~(c).
Each colored region denotes the spatial extent of a detector of a particular basis, with time progressing top-to-bottom, left-to-right.

The most important feature of these diagrams is the shear structure we observe in the detector slices for the measurement-maximized schedule, that we do not observe in the vanilla schedule.
This is produced by two diagonals of shapes that open towards each other (as in Figure~\ref{fig:stretch}), increasing the support of operators between the diagonals, while decreasing the support of shapes directly outside the diagonals.
Although it may not be immediately obvious why such a structure is undesirable, we can make a simple path-counting argument to explain.

We consider the graph in which vertices correspond to qubits and two vertices are connected by an edge if the corresponding qubits are in the support of the same detector.
Although we can consider this graph for the entire set of detectors at once, for this argument it is enough to consider the end-cycle state.
Then paths through this graph correspond to logical errors.
The length of the shortest such path is the exponent of the leading order term of the probability of logical error, while the number of minimum weight paths is the coefficient of this term.
The circuit distance of the circuits corresponding to the two plays are identical, meaning that the leading order is the same in each case.
The number of paths, however, is different.

If we consider a path leading from the left side to the right side of the end-cycle state, we can see that this shear structure creates more minimum weight paths.
From each of the large detectors, there are three choices to continue from left to right -- two to another large detector, and one to a small detector.
This is in contrast to the case where the detectors are more evenly sized -- there are almost no detectors from which one can make more than two choices to continue left to right.

On top of the increase in number of space-like paths per time-slice, the schedule corresponding to maximizing measurements creates a less balanced set of detectors in
\clearpage
\noindent
time; that is, there are more of the shortest-in-time detectors as well as more of the longest-in-time.
This also creates more minimum weight paths through the graph.
In theory, optimizing to reduce the number of minimum weight paths directly could decrease LER further, but such a condition is difficult to encode in an ILP.
\section{Comparison to ACID}\label{appendix:acid}
Given the shared optimization goals, we benchmark the LER performance of surface code measurement schedules generated by ACID~\cite{wolanski2025automatedcompilationincludingdropouts} against the canonical implementation of vanilla LUCI.
To ensure a consistent comparison under the noise model used throughout this work, we adapted the ACID schedules into LUCI plays prior to simulation. 
This adaptation was necessary for two reasons:
\begin{enumerate}
    \item It aligns the noise model by replacing the noiseless state preparation and readout (Stim's MPP) assumed in ACID with the noisy operations used in our baseline when compiling the play to a circuit.
    \item It ensures the resulting detectors are graphlike, enabling the use of the same matching-based decoder for both datasets to isolate schedule performance from decoder capability.
\end{enumerate}

We also slightly modify some boundary conditions.
For some configurations of dropout, ACID produces a circuit which measures a set of operators that do not allow one to find a set of superstabilizers and gauge operators with canonical (anti)commutation relations, (e.g. producing $X_1X_2X_3, X_4X_5X_6, Z_3Z_4Z_7$, without including $Z_1Z_6Z_8$ since qubit $8$ is outside the boundaries of the surface code patch).
When this occurs, we remove unpairable operators (without decreasing circuit distance) until the expected gauge group can be found.
This behavior only occurs when a dropout region touches the boundary.

\begin{figure}[H]
    \centering
    \begin{subfigure}[b]{\linewidth}
        \centering
        \includegraphics[width=\linewidth]{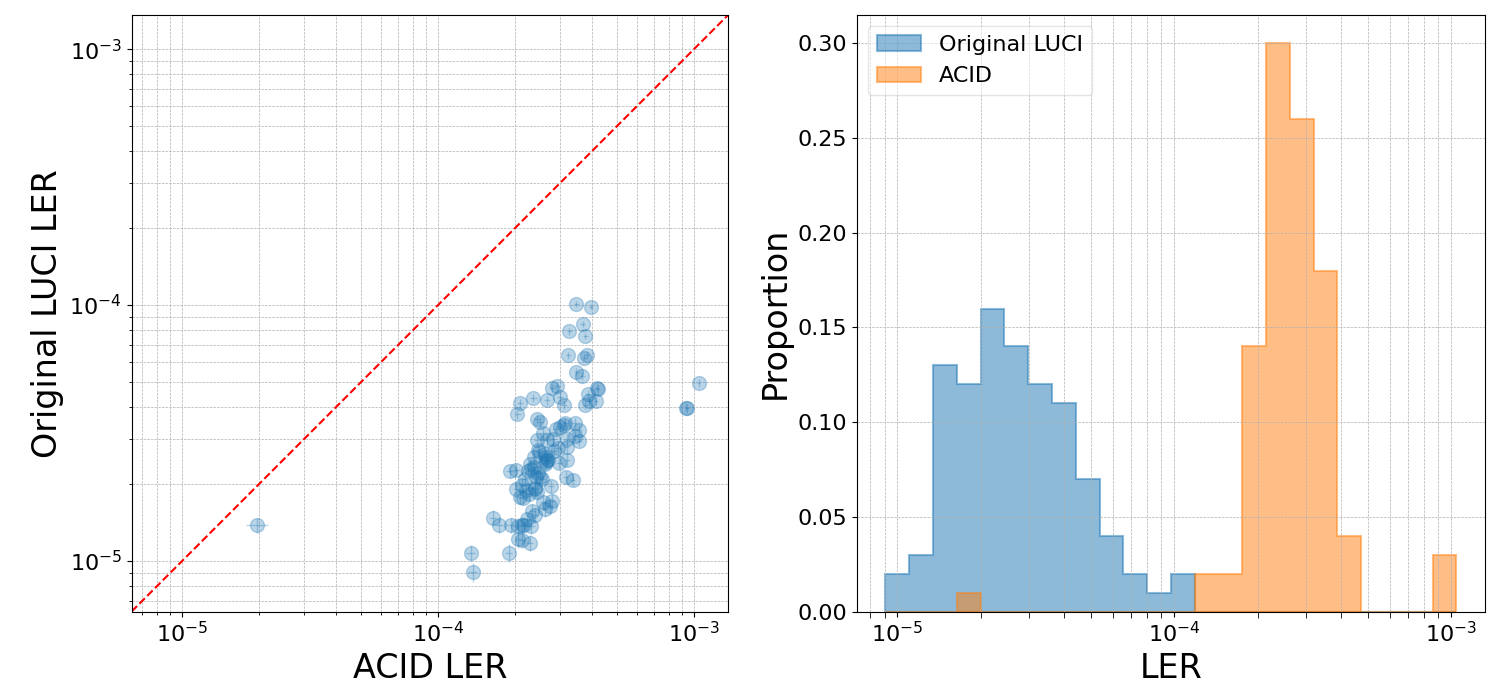} 
        \caption{LER comparison at $1\%$ dropout.}
    \end{subfigure}
    \begin{subfigure}[b]{\linewidth}
        \centering
        \includegraphics[width=\linewidth]{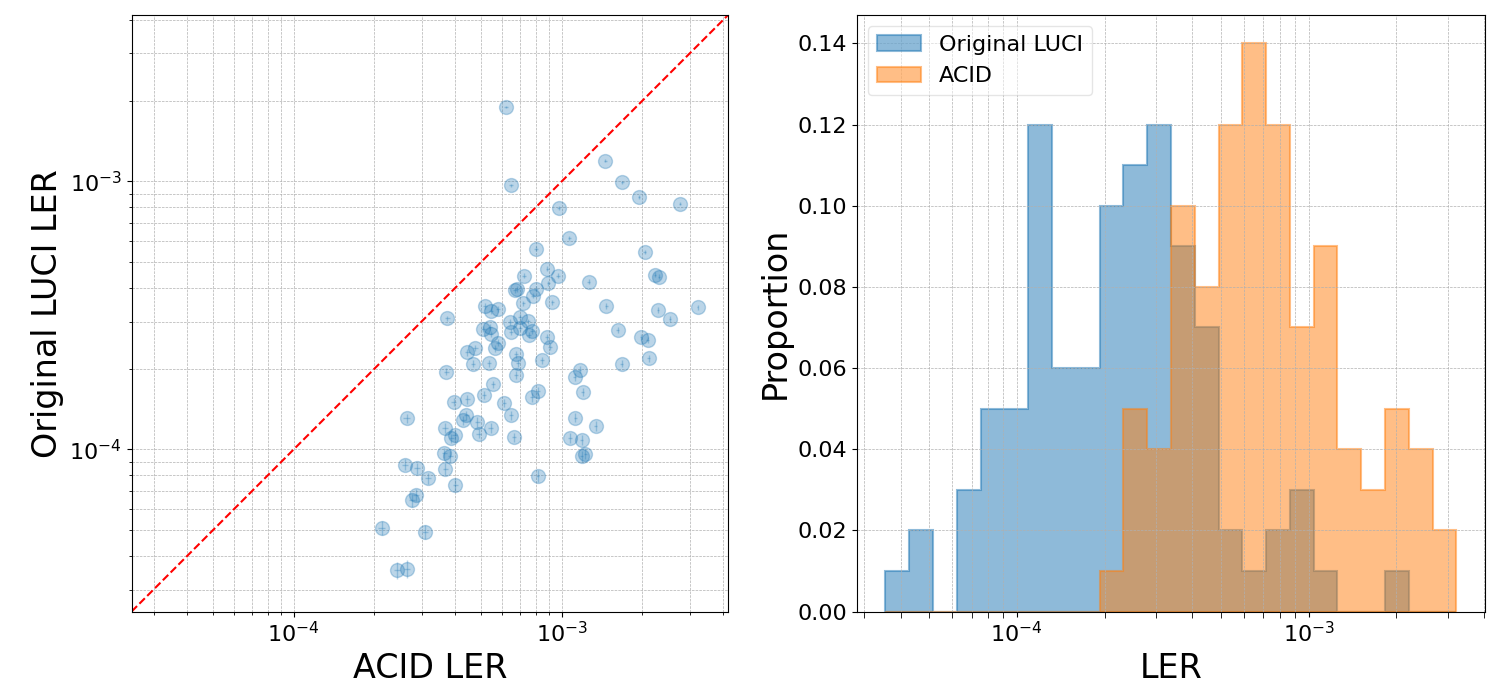}
        \caption{LER comparison at $3\%$ dropout.}
    \end{subfigure}
    \caption{LER for ACID compared to LER for our implementation of LUCI.}
    \label{fig:ACID_vs_LUCI}
\end{figure}

In Figure~\ref{fig:ACID_vs_LUCI}, we compare the performance of these approaches on the fixed ensemble of dropout configurations we have used throughout, simulating performance under the same noise model at $0.1\%$. 
We take the approach given by ACID, performing $d$ blocks, each block consisting either of $2, 3$ or $4$ rounds, depending on the length of the schedule found (with LUCI always producing a length $4$ schedule).
It should be noted that although in the original paper, ACID was able to find a three-round measurement schedule in all considered cases, the higher dropout rates that we consider here mean that some ACID schedules are also four rounds (for instance, two adjacent broken couplers forming a line require a four round solution).

We observe that the canonical vanilla LUCI implementation yields consistently lower logical error rates than ACID schedules.
We attribute the difference in observed LER for LUCI to the difficulty of fully reconstructing the vanilla LUCI logic solely from literature. 
The comparative baseline used in \cite{wolanski2025automatedcompilationincludingdropouts} maximizes measurement counts while ensuring operators are measured at least once every four rounds.
However, the canonical implementation includes subtle geometric assumptions -- specifically regarding shape orientation -- that are critical but not immediately obvious.
With these orientations, the canonical implementation largely avoids the large, unbalanced detecting regions that appear when the primary objective is simply maximizing measurements.

Maximizing measurements only is similar to setting $\beta, \gamma,\delta = 0$ in our optimized implementation, which we show the results of in Figure~\ref{fig:max_vs_orig}.
The fact that our model slightly outperforms ACID in this case as well is due to the fact that we `hint' the solver to use solutions similar the default algorithm when the objective function does not prefer one solution over another, as well as the fact that three-round solutions generally produce a worse LER than four-round solutions (see Appendix~\ref{appendix:3_round}).

We hope that incorporating heuristics such as those discussed in Section~\ref{sec:ilp-model} can also help to improve LER for those applications of ACID to which LUCI does not apply, i.e. the superdense color code~\cite{gidney2023newcircuitsopensource} and the bivariate bicycle code~\cite{bravyi2024bb}.
\end{document}